\documentclass[preprint,12pt,authoryear]{elsarticle}

\usepackage{amssymb}
\usepackage{hyperref}

\journal{New Astronomy Reviews}

\begin{document}
\newcommand\aap{A\&A}                % Astronomy and Astrophysics
\let\astap=\aap                          % alternative shortcut
\newcommand\aapr{A\&ARv}             % Astronomy and Astrophysics Review (the)
\newcommand\aaps{A\&AS}              % Astronomy and Astrophysics Supplement Series
\newcommand\actaa{Acta Astron.}      % Acta Astronomica
\newcommand\afz{Afz}                 % Astrofizika
\newcommand\aj{AJ}                   % Astronomical Journal (the)
\newcommand\ao{Appl. Opt.}           % Applied Optics
\let\applopt=\ao                         % alternative shortcut
\newcommand\aplett{Astrophys.~Lett.} % Astrophysics Letters
\newcommand\apj{ApJ}                 % Astrophysical Journal
\newcommand\apjl{ApJ}                % Astrophysical Journal, Letters
\let\apjlett=\apjl                       % alternative shortcut
\newcommand\apjs{ApJS}               % Astrophysical Journal, Supplement
\let\apjsupp=\apjs                       % alternative shortcut
% The following journal does not appear to exist! Disabled.
%\newcommand\apspr{Astrophys.~Space~Phys.~Res.} % Astrophysics Space Physics Research
\newcommand\apss{Ap\&SS}             % Astrophysics and Space Science
\newcommand\araa{ARA\&A}             % Annual Review of Astronomy and Astrophysics
\newcommand\arep{Astron. Rep.}       % Astronomy Reports
\newcommand\aspc{ASP Conf. Ser.}     % ASP Conference Series
\newcommand\azh{Azh}                 % Astronomicheskii Zhurnal
\newcommand\baas{BAAS}               % Bulletin of the American Astronomical Society
\newcommand\bac{Bull. Astron. Inst. Czechoslovakia} % Bulletin of the Astronomical Institutes of Czechoslovakia 
\newcommand\bain{Bull. Astron. Inst. Netherlands} % Bulletin Astronomical Institute of the Netherlands
\newcommand\caa{Chinese Astron. Astrophys.} % Chinese Astronomy and Astrophysics
\newcommand\cjaa{Chinese J.~Astron. Astrophys.} % Chinese Journal of Astronomy and Astrophysics
\newcommand\fcp{Fundamentals Cosmic Phys.}  % Fundamentals of Cosmic Physics
\newcommand\gca{Geochimica Cosmochimica Acta}   % Geochimica Cosmochimica Acta
\newcommand\grl{Geophys. Res. Lett.} % Geophysics Research Letters
\newcommand\iaucirc{IAU~Circ.}       % IAU Cirulars
\newcommand\icarus{Icarus}           % Icarus
\newcommand\japa{J.~Astrophys. Astron.} % Journal of Astrophysics and Astronomy
\newcommand\jcap{J.~Cosmology Astropart. Phys.} % Journal of Cosmology and Astroparticle Physics
\newcommand\jcp{J.~Chem.~Phys.}      % Journal of Chemical Physics
\newcommand\jgr{J.~Geophys.~Res.}    % Journal of Geophysics Research
\newcommand\jqsrt{J.~Quant. Spectrosc. Radiative Transfer} % Journal of Quantitiative Spectroscopy and Radiative Transfer
\newcommand\jrasc{J.~R.~Astron. Soc. Canada} % Journal of the RAS of Canada
\newcommand\memras{Mem.~RAS}         % Memoirs of the RAS
\newcommand\memsai{Mem. Soc. Astron. Italiana} % Memoire della Societa Astronomica Italiana
\newcommand\mnassa{MNASSA}           % Monthly Notes of the Astronomical Society of Southern Africa
\newcommand\mnras{MNRAS}             % Monthly Notices of the Royal Astronomical Society
\newcommand\na{New~Astron.}          % New Astronomy
\newcommand\nar{New~Astron.~Rev.}    % New Astronomy Review
\newcommand\nat{Nature}              % Nature
\newcommand\nphysa{Nuclear Phys.~A}  % Nuclear Physics A
\newcommand\pra{Phys. Rev.~A}        % Physical Review A: General Physics
\newcommand\prb{Phys. Rev.~B}        % Physical Review B: Solid State
\newcommand\prc{Phys. Rev.~C}        % Physical Review C
\newcommand\prd{Phys. Rev.~D}        % Physical Review D
\newcommand\pre{Phys. Rev.~E}        % Physical Review E
\newcommand\prl{Phys. Rev.~Lett.}    % Physical Review Letters
\newcommand\pasa{Publ. Astron. Soc. Australia}  % Publications of the Astronomical Society of Australia
\newcommand\pasp{PASP}               % Publications of the Astronomical Society of the Pacific
\newcommand\pasj{PASJ}               % Publications of the Astronomical Society of Japan
\newcommand\physrep{Phys.~Rep.}      % Physics Reports
\newcommand\physscr{Phys.~Scr.}      % Physica Scripta
\newcommand\planss{Planet. Space~Sci.} % Planetary Space Science
\newcommand\procspie{Proc.~SPIE}     % Proceedings of the Society of Photo-Optical Instrumentation Engineers
\newcommand\rmxaa{Rev. Mex. Astron. Astrofis.} % Revista Mexicana de Astronomia y Astrofisica
\newcommand\qjras{QJRAS}             % Quarterly Journal of the RAS
\newcommand\sci{Science}             % Science
\newcommand\skytel{Sky \& Telesc.}   % Sky and Telescope
\newcommand\solphys{Sol.~Phys.}      % Solar Physics
\newcommand\sovast{Soviet~Ast.}      % Soviet Astronomy (aka Astronomy Reports)
\newcommand\ssr{Space Sci. Rev.}     % Space Science Reviews
\newcommand\zap{Z.~Astrophys.}       % Zeitschrift fuer Astrophysik

\begin{frontmatter}



\title{Relativistic Jets of Blazars}


\author[label1,label2]{Talvikki Hovatta}
\author[label1]{Elina Lindfors}

\address[label1]{Finnish Centre for Astronomy with ESO (FINCA), University of Turku, FI-20014 University of Turku, Finland}
\address[label2]{Aalto University Mets\"ahovi Radio Observatory, Mets\"ahovintie 114, FI-02540 Kylm\"al\"a, Finland}

\begin{abstract}
Relativistic jets of active galactic nuclei have been known to exist for 100 years. Blazars with their jet pointing close to our line of sight are some of the most variable and extreme objects in the universe, showing emission from radio to very-high-energy gamma rays. In this review, we cover relativistic jets of blazars from an observational perspective with the main goal of discussing how observations can be used to constrain theoretical models. We cover a range of topics from multiwavelength observations to imaging of jets with a special emphasis on current open questions in the field.

\end{abstract}


\end{frontmatter}


\section{Introduction}\label{intro}

The term blazar was first suggested by Ed Spiegel 40 years ago in a conference dinner talk at the Pittsburgh Conference on BL~Lac Objects \citep{wolfe78}. The term was introduced to find a common name for BL~Lac objects and flat spectrum radio quasars (FSRQs), which back then were mostly called optically violently variable quasars. Blazars are active galactic nuclei (AGN) that host relativistic jets that are pointing very close to our line of sight. This results in their emission being highly beamed and Doppler boosted, making them bright and variable in all wavebands from radio to $\gamma$-rays.

The spectral energy distributions (SEDs) of both subclasses of blazars show two peaks, where the first peak is generally attributed to synchrotron emission. The peak frequency of the synchrotron bump is commonly used to further divide the blazars into  low-,  intermediate-  and  high  frequency  peaked sources  (LSP,  ISP  and  HSP,  respectively)  with  log $\nu_{peak}<14$ defining the LSP, $14<$log $\nu_{peak}<15$ the ISP, and log $\nu_{peak}>15$ the HSP classes \citep{abdo10c}. In the case of FSRQs, the first SED peak is typically in the infrared regime (i.e. they are LSPs) while for BL~Lacs it can be anywhere between infrared and hard X-rays (LSPs, ISPs and HSPs).

The second SED peak is generally attributed to inverse Compton scattering. Seed photons for the inverse Compton scattering are provided by the synchrotron emission \citep{maraschi92} or can originate from the dense radiation field generated by the direct and reprocessed accretion disk emission \citep{dermer93,sikora94} or molecular torus \citep{sikora08}. Also hadronic mechanisms for producing the second peak have been proposed \citep{mannheim93,mucke01}. The optical spectra of FSRQs show broad emission lines, implying an existence of fast moving gas clouds close (0.1 to 1 parsec) to the central engine, while BL~Lacs show very weak or no emission lines in their spectra. Therefore, in general, synchrotron self Compton (SSC) models are favored for BL~Lac objects and external Compton (EC) models for FSRQs.

The detection of high-energy gamma-rays from Narrow-Line Seyfert 1 Galaxies (NLS1) \citep{abdo09a,abdo09b,abdo09c} suggests that also they host relativistic jet pointing close to our line of sight and indeed superluminal motion has been detected in some NLS1 ~\citep{lister16}. This would qualify them as blazars. However, in this review, we do not discuss them, but point the interested readers to \citet{foschini17,Aaltodoc:http://urn.fi/URN:ISBN:978-952-60-8154-0}, and references therein.

In this review, we have mainly concentrated on the observational properties of blazars at different wavelengths, and how the observations can be used to constrain theoretical models. In Section~\ref{obstools}, we cover the different observational tools that can be used to study blazars by utilizing their SEDs, light curves and spatially resolved very long baseline interferometry (VLBI) observations. In Section~\ref{questions}, we have listed (in our opinion) the most important open questions that still need further understanding. This section also includes details on SED modeling (Section~\ref{sedparameters}), magnetic fields (Section~\ref{Bfields}), and structure of the jets (Section~\ref{structure}). We end this review with an outlook in Section~\ref{outlook}, where we have listed some major upcoming new instruments that will help to shed light on the open questions.

In the last years, there have also been other reviews on relativistic jets in AGN and blazars. \cite{blandford18} includes a very comprehensive review on relativistic jets in AGN, covering a broader range of topics, such as the unification of radio galaxies and blazars. Their review also includes a more in-depth discussion on the emission processes and theoretical models for AGN emission. \cite{bottcher19} also covers recent progress made in multimessenger observations and the theory of blazars, with emphasis on theoretical models, and, for example, leaving radio emission out of the picture. We consider both of these reviews to be complementary to this review, where we have covered many aspects of blazars from the viewpoint on how observations can be used to answer the open questions.

\section{Observational tools}\label{obstools}
\subsection{Multiwavelength observations}

As blazars are bright and variable in all bands from radio to very high energy (VHE) $\gamma$-rays, coordinated multiwavelength observations have been the key tool for blazar studies. In the 1990s, with the launch of the Compton $\gamma$-ray observatory and Rossi X-ray Timing Explorer telescope, the first campaigns including X-ray and $\gamma$-ray observations were organized. The campaigns typically consisted of quasi-simultaneous data from radio, near-IR, optical, X-rays and $\gamma$-rays. A worldwide network of optical, near-IR and radio observers, (whole earth blazar telescope, WEBT, \citealt{villata02}) was established to organize and support such campaigns. The multiwavelength campaigns typically had duration of weeks to months. These observations revealed that in many sources the total energy ($\nu$$F_\nu$) radiated by blazars is largely dominated by the $\gamma$-ray band, in other words, the sources were found to be strongly Compton dominated \citep{hartman92}. 

\begin{figure}
\includegraphics[scale=0.66]{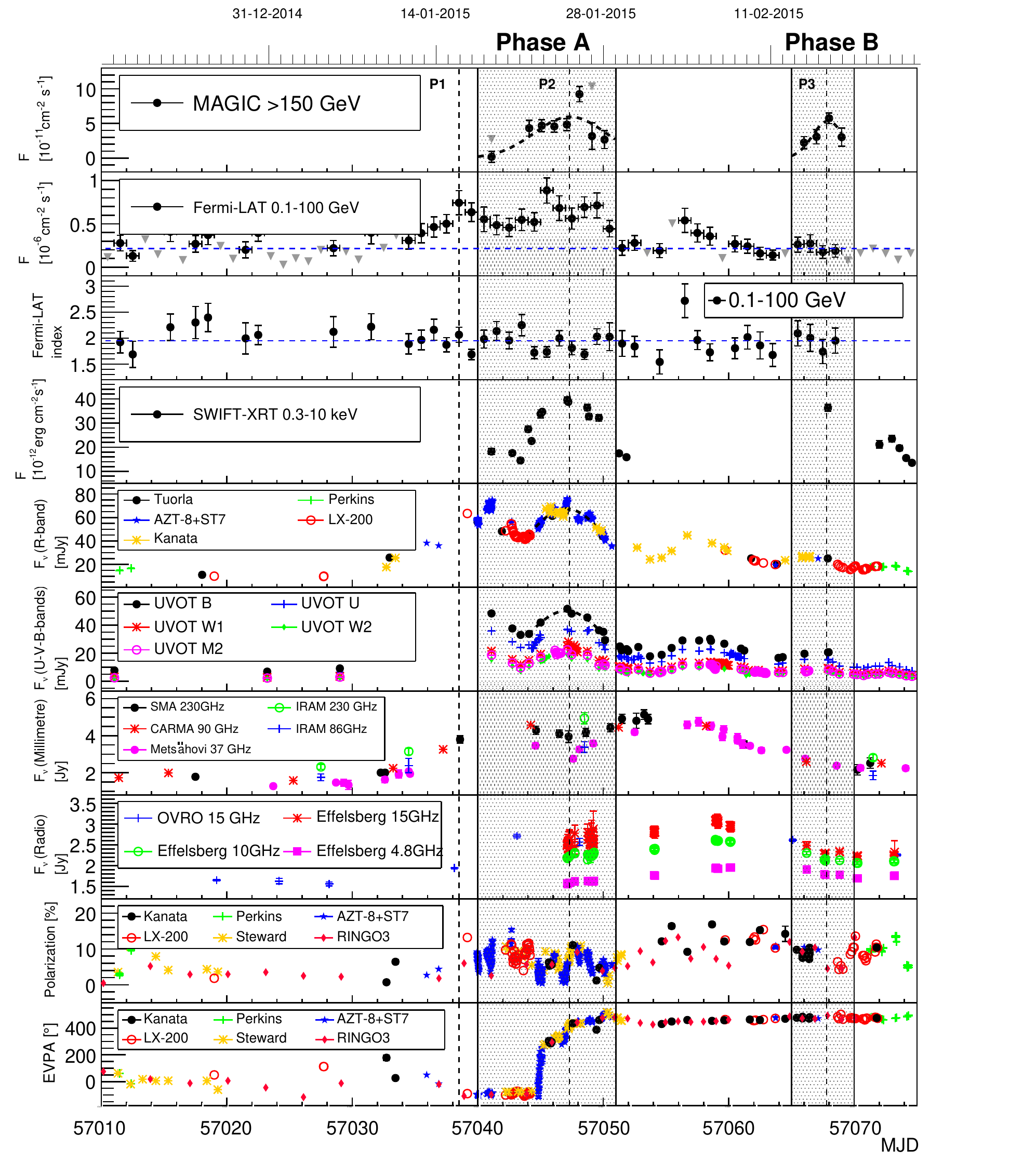}
\caption{Example of a recent extensive multiwavelength data set including data from radio to VHE $\gamma$-rays. The source is S5~0716+714 and demonstrates the complexity of the flaring behavior of blazars: the phase A shows correlated variability in all bands and a fast rotation of electric vector position angle (EVPA), while the phase B flare is only visible in VHE $\gamma$-rays and X-rays. The spectral energy distribution and VLBA data for the same campaign are shown in Figure~\ref{fig:0716b}. Figure adapted from \cite{ahnen18a}. Reproduced with permission from Astronomy \& Astrophysics, \copyright ESO.}
%[Needs permission from A\&A]}
\label{fig:0716a}
\end{figure}

\begin{figure}
\includegraphics[scale=0.66]{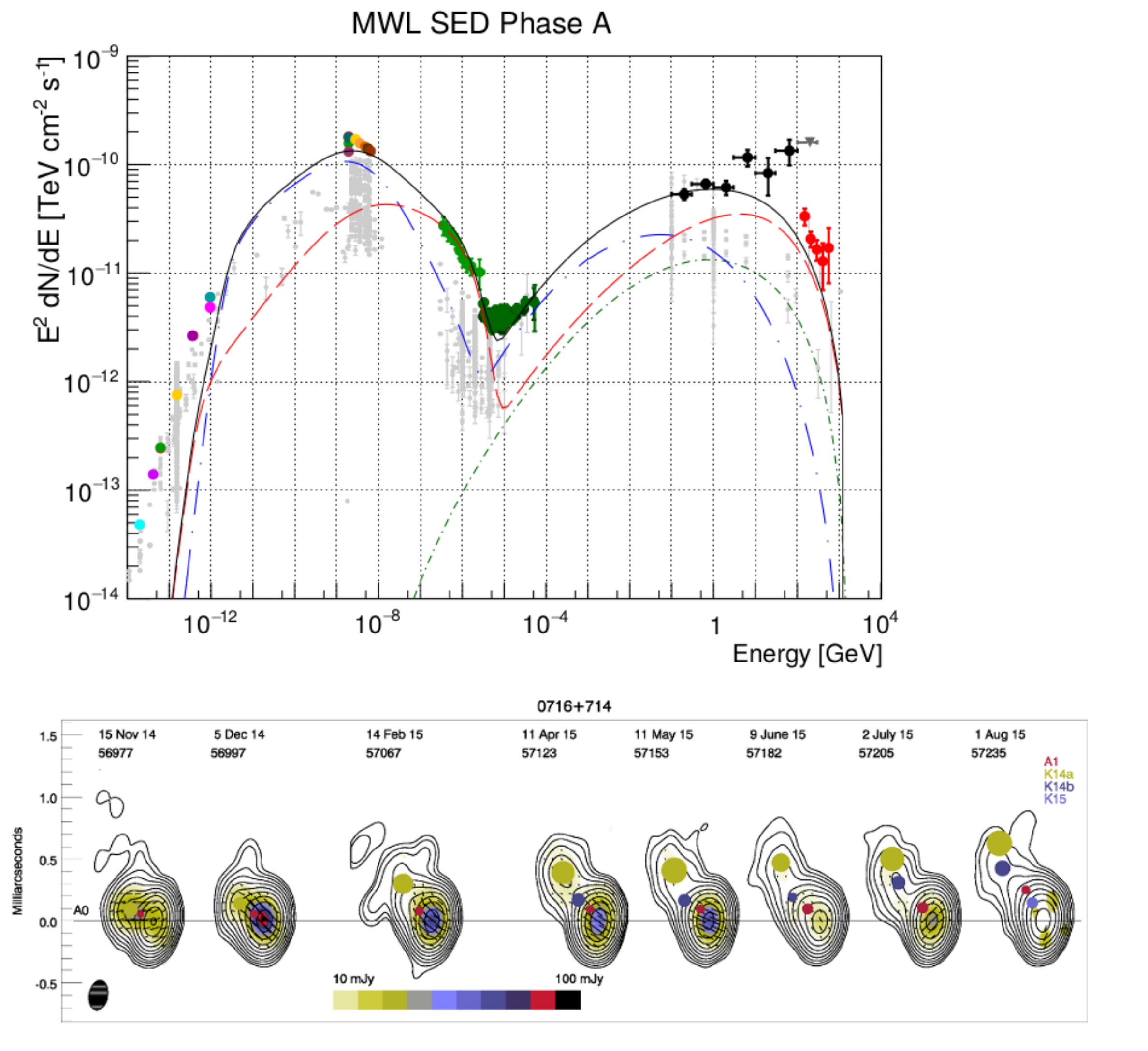}
\caption{Example of a recent extensive multiwavelength data set including data from radio to VHE $\gamma$-rays: SED on the top, VLBA data in the bottom. The source is S5~0716+714 and the SED shows the wide separation of the two SED peaks: the synchrotron peak is in the optical range, the IC peak at several GeV. The dark green data from NuStar constrains the transition between these two peaks very accurately. The VLBA data (lower panel) from the Boston VLBA blazar monitoring program at 43\,GHz shows a new component emerging from the VLBA core around the high activity period. Figure adapted from \cite{ahnen18a}. Reproduced with permission from Astronomy \& Astrophysics, \copyright ESO.}
\label{fig:0716b}
\end{figure}

Due to the limited sensitivity of the EGRET instrument, the campaigns were limited to a handful of the brightest blazars, and only in few cases variability in the $\gamma$-ray band could be studied in connection with variability in other bands. This has been revolutionized by the {\it Fermi}-LAT $\gamma$-ray satellite, which has detected more than 1500 blazars in $\gamma$-ray band (1591 in 3LAC, \citealt{ackermann15a}). For a comprehensive review on the {\it Fermi} $\gamma$-ray sky and the multiwavelength connection see \cite{massaro15}. Well-sampled, quasi-simultaneous data sets have now become available for a larger number of sources, and the data sets also include VHE $\gamma$-ray data. While the first VHE $\gamma$-ray emitting blazar, Mrk~421, was discovered already in 1991 \citep{punch92}, only six VHE $\gamma$-ray emitting blazars had been detected by 2000. The current generation of Imaging Air Cherenkov Telescopes (IACTs), H.E.S.S., MAGIC and VERITAS, have increased the number of VHE $\gamma$-ray blazars to $\sim80${\footnote{\url{http://tevcat.uchicago.edu/}}}, now also covering all blazar classes, even if HSPs are still by far the most numerous. Figures~\ref{fig:0716a} and \ref{fig:0716b} demonstrate one of the recent extensive multiwavelength campaigns with excellent coverage from radio to VHE $\gamma$-rays.

In the following sections, we describe the major advances in multiwavelength observations of blazars, first in relation to spectral energy distributions and then to light curves.

\subsubsection{Spectral energy distributions} \label{sed}

Spectral energy distributions (SEDs) have been the main tools for blazar studies since the early days, and extensive multiwavelength campaigns have been organized in order to have simultaneous data from radio to $\gamma$-ray bands \citep[see][for the most extensive campaigns in the CGRO era]{hartman01}. Since the launch of {\it Fermi}-LAT, several campaigns have been organized, the most persistent being the campaigns on Mrk~421 and Mrk~501 \citep{abdo11a,abdo11b} including all main instruments. It is, however, also very important to study the SEDs of larger samples of blazars as done in e.g., \cite{giommi12a}. Nowadays, a large sample of SEDs is collected in the Space Science Data Center{\footnote{\url{https://tools.ssdc.asi.it/}}}.
As the availability of the data has been improving, it has also become more and more evident that single snapshot SEDs cannot constrain the emission models sufficiently, and they have to be combined with other observations. This will be discussed in section \ref{sedparameters}.

In recent years, daily "time-lapses" of the evolution of the SED have become possible \citep{aleksic15,krauss16}. This also requires new approaches to the modeling of the SEDs, and instead of snapshot single-epoch models, the modeling should be performed self-consistently including time-evolution. This also provides a new way to constrain the physics of the whole jet more consistently \citep{lucchini19}. 

\subsubsection{Intra-day light curves} \label{intraday}

In the past years, fast variability of blazars in VHE $\gamma$-rays has gained a lot of attention. It was first seen from the brightest VHE blazars, Mrk~421, PKS~2155-304, and Mrk~501 \citep{gaidos96,aharonian07,albert07}, but has also now been detected from FSRQs \citep{aleksic11} and BL~Lac \citep{arlen13} (which is a borderline ISP/LSP source).
The variability time scales seen in VHE $\gamma$-rays are shorter than (or on the order of) 10 minutes. These are the shortest variability time scales that have been detected from blazars, due to the much smaller collecting area of {\it Fermi} (versus the IACTs), such fast variability has been detected only from 3C~279 in the {\it Fermi} band \citep{ackermann16}.

It should be noted, that in the lower energy bands of X-ray, optical and radio, intra-day observations have been performed already for decades and short-term variability was detected, for example, in S5~0716+714 \citep{wagner96}. The amplitude of the variability was smaller, but it was still clear that explaining this variability requires extreme physical conditions. 

Simultaneous VHE-optical and $\gamma$-optical intra-day light curves, covering the phase of fast variability in the highest energies are still extremely rare. For PKS~1510$-$089, such light curves were obtained, and there seems to be similar patterns in all three (VHE, GeV and optical) light curves, even though the variability amplitudes are different \citep{zacharias17}. 

The observed variability time scale imposes a limit to the emission region size, so the emission region from which the sub-hour variability originates from must be small. At the same time, there is no evidence for internal  $\gamma-\gamma$-absorption by the co-spatially produced low-energy (IR -- X-ray) radiation in the observed $\gamma$-ray emission of blazars, so the emission region must be optically thin to this process. This implies that the Doppler factor must be high $\delta\sim50$, much higher than the Doppler factors typically observed in the jets with VLBI observations. In addition, it is extremely difficult to accelerate particles in so short time scales. This problem has gained a lot of attention in past years, which is nicely summarized in a dedicated section in the recent review by \citet{bottcher19}, in our review we discuss this issue only shortly in Sections~\ref{coreblobs} and \ref{seedphotons}.

\subsubsection{Long-term light curves} \label{longterm}

Already in the late 1980's there were some blazars for which it was possible to construct 100-year long optical light curves by extracting data from old photographic plates, most notably OJ~287 for which a 12-year periodicity was suggested based on nearly 100 years of data \citep{sillanpaa88}. There are few other sources for which several tens of years of optical data exists such as 3C~273 \citep{smith63} and BL~Lacertae \citep{villata04}. As remotely operated and robotic telescopes have become more common in last 15 years, also the number of sources for which long-term optical data exists have become more numerous, and nowadays there are several on-going optical monitoring programs that have already collected more than 10 years of data  \citep{nilsson18,smith09}. 

In the radio wavelengths, the earliest monitoring programs, such as the University of Michigan Radio Astronomy Observatory (UMRAO), Mets\"ahovi Radio Observatory and RATAN-600 blazar monitoring programs began in the 1960s-70s \citep{aller85,salonen87,korolkov79}. The UMRAO program, operating at three cm-band frequencies (4.8, 8, and 14.5\,GHz) continued until 2012, providing more than 40 years of radio data both in total intensity and polarization. At Mets\"ahovi, the monitoring is still on-going, and is concentrated on shorter mm-band wavelengths, being mainly done at a frequency of 37\,GHz, corresponding to 8\,mm. RATAN-600 is unique in that it can provide simultaneous observations at 1.1, 2.3, 4.8, 7.7, 11.2, and 21.7\,GHz frequencies, which is especially useful when studying the shape of the radio spectrum \citep[e.g.,][]{mingaliev14}.

\begin{figure}
\includegraphics[width=\hsize]{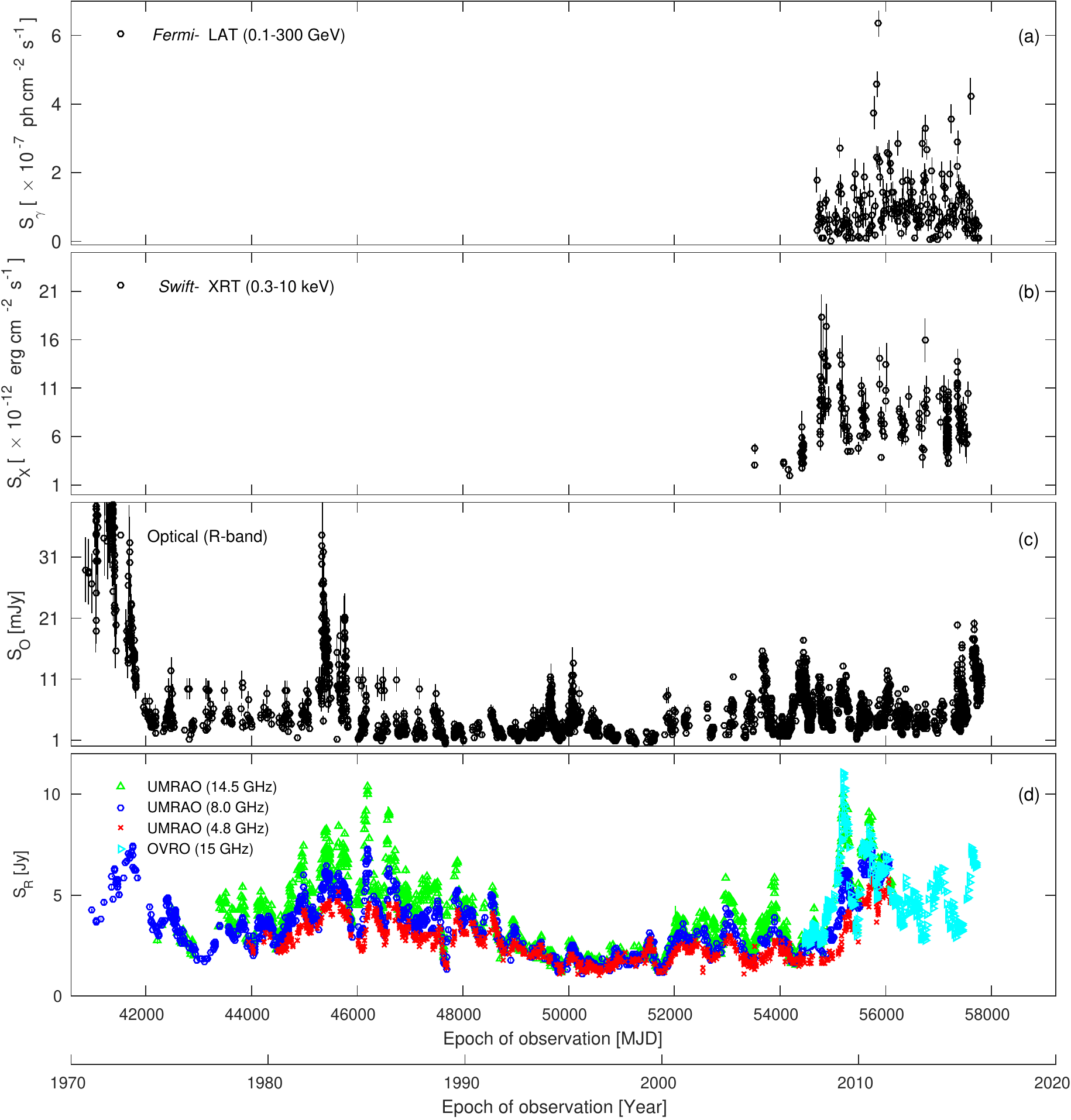}
\caption{From top to bottom: Long-term light curves of OJ~287 in $\gamma$-rays observed by {\it Fermi}-LAT, X-rays by {\it Swift}-XRT, optical R-band by multiple observatories, and 4.8, 8, and 15\,GHz radio by UMRAO and OVRO monitoring programs. Figure taken from \cite{goyal18}, where the time series were analyzed using multiple statistical methods. Reproduced by permission of the AAS.}
\label{fig:goyal}
\end{figure}

In the last decade, to especially support the {\it Fermi} $\gamma$-ray mission there have been more radio monitoring programs, such as the Owens Valley Radio Observatory (OVRO) 40-m monitoring program, which has obtained 15\,GHz light curves of nearly 2000 sources since 2008 \citep{richards11}. The F-GAMMA program monitored over 60 blazars at multiple frequencies from 2 to 345\,GHz to uncover spectral changes and frequency dependencies in the variability \citep{fuhrmann14,fuhrmann16}. At shorter mm-wavelengths, the POLAMI program \citep{agudo18} has followed a smaller number of sources at 3 and 1.3\,mm wavelengths.

In X-rays {\it Rossi} X-ray timing explorer (RXTE) enabled the first long-term monitoring observations of a handful of the strongest X-ray blazars. For example, \citet{chatterjee08} analyzed more than 10 years of X-ray variability of 3C~279 together with radio and optical data. The X-ray all sky monitors such as ASM (onboard of RXTE) and MAXI-GSC are limited to monitoring the brightest X-ray blazars, such as Mrk~421 \citep[e.g.][]{isobe15}, but are still ultimately the only way to get X-ray light curves with good cadence. While {\it Swift} has been crucial for blazar SED studies \citep[see][for a review]{ghisellini15}, the long-term light curves are rather sparsely sampled and most of the blazars have been observed mainly during the multiwavelength campaigns triggered by flares, again with the exception of Markarian 421 \citep{carnerero17}.

While EGRET was in operation for almost 10 years, its pointing mode observations and limited sensitivity did not really allow producing long-term light curves of blazars in $\gamma$-rays. The situation changed dramatically with {\it Fermi}-LAT, which has been scanning the $\gamma$-ray sky continuously, every 3 hours, for already more than a decade. Still, long-term $\gamma$-ray light curves have been analyzed in detail only for rather small samples of sources \citep[e.g. 13 sources in][]{sobolewska14}. In Fig.~\ref{fig:goyal}, an example of long-term radio, optical, X-ray, and $\gamma$-ray light curves are shown for OJ~287. Variability is clearly seen in all wavebands. Notable are the much longer light curves available for the radio and optical bands.  

Also in VHE $\gamma$-rays the real long-term light curves are limited to a handful of the brightest sources in that energy band, in particular Markarian 421 \citep{magic_mrk421,veritas_mrk421} and PKS~2155$-$304 \citep{hess17}.
While the number of blazars detected in VHE $\gamma$-ray blazars has been steadily growing with the advent of H.E.S.S., MAGIC and VERITAS telescopes, the VHE $\gamma$-ray observations are often triggered by flares in the other wavebands, and most of the blazars are mainly observed during flares. Therefore, dedicated, but poorer sensitivity monitoring telescopes Whipple\footnote{\url{https://veritas.sao.arizona.edu/whipple-10m-topmenu-117}} and FACT \citep{dorner17} are an important addition when studying the long-term behavior of blazars in VHE $\gamma$-rays. 
%VHE gamma-rays, Markarians, FACT, \citep{hess17}

With the growing amount of data in the last decades, and especially with the upcoming wide-field instruments that are expected to transfer the field into the direction of big data, it is necessary to develop and use statistical tools to study blazars. Moreover, for understanding the big picture, and the connection between different blazar types, large samples of objects need to be studied objectively. First studies of the long-term data sets included typically the use of structure functions \citep{simonetti85} to analyze the characteristic time scales in the radio light curves \citep{hughes92,hufnagel92,lainela93}. 

Other often-used methods included correlation functions to study the connection between different wavebands, typically between optical and radio, which had the best-sampled data trains \citep{hufnagel92, tornikoski94,clements95,hanski02}. Especially the Discrete Correlation Function (DCF) developed to work on irregularly sampled data \citep{edelson88,hufnagel92} was found to be ideal for studying both the characteristic time scales of sources by the use of autocorrelation \citep[e.g.,][]{hovatta07}, and the correlations between two wavebands. Additionally, different types of periodograms, such as the Lomb-Scargle periodogram \citep{lomb76,scargle82}, were developed for investigating possible periodicities in the light curves (see also Section \ref{periodicity}).

In the last decade, the focus has shifted more towards Fourier-based methods, such as Power Spectral Density (PSD) analysis, which are used to study both the underlying physical process behind the variations, and possible periodicities in the light curves. The PSDs of blazars are most often modeled by a single power-law function, where the spectral slope indicates the nature of the variations. For example, slope of 0 corresponds to white noise where the variations at different frequencies (or time scales) are not correlated, a slope of 1 indicates pink noise, and a slope of 2 a red noise process. The steepness of the slope describes the relative contribution of different frequencies to the variability, with a steeper slope indicating that long time scales dominate the variability. In the radio band, the PSD slope is typically around 2 \citep[e.g.,][]{max-moerbeck14,park14,ramakrishnan15}, while in the optical and $\gamma$-ray bands it is around 1.5 \citep[e.g.,][]{abdo10b,chatterjee12,max-moerbeck14,nilsson18}, indicating that in the radio band, longer time scales dominate, in comparison to higher energies. Spectral breaks in the PSD would indicate the presence of characteristic time scales or periodicities in the light curves. These will be discussed in Section \ref{periodicity}.

The combined use of PSD analysis to uncover the underlying stochastic process behind the variations, and DCF has allowed to study the connection between variations in radio, optical and $\gamma$-ray bands \citep[e.g.,][]{max-moerbeck14, cohen14, ramakrishnan15,ramakrishnan16}. These studies show that the radio flares typically lag the $\gamma$-ray flares by some tens to hundreds of days, with shorter delays at higher frequencies \citep{fuhrmann14, ramakrishnan16}. In the optical, both simultaneous and short delays with respect to $\gamma$-ray variability is seen \citep{cohen14, ramakrishnan16}.

Recently, \cite{goyal18} studied the variability of the blazar OJ~287 at multiple bands from radio to $\gamma$-ray energies (see Fig.~\ref{fig:goyal} for the long-term light curves used in the study) using the continuous-time autoregressive moving average (CARMA) model by \cite{kelly14}. In this case, the PSDs are first calculated in the time domain, which makes treating unevenly sampled data easier. Their optical data set included observations from the Kepler satellite, allowing them to probe the time scales over six decades in frequency. They found that in the radio, optical and X-ray bands, the variability is well characterized by colored noise from years down to minutes time scales, while in the $\gamma$-rays the variability at shorter than 150 days time scales seems to be dominated by uncorrelated white noise. Also, other likelihood-based methods such as modeling the light curves using Ornstein-Uhlenbeck (OU) process \citep{kelly09, kelly11} have been used to estimate the variability characteristics of {\it Fermi}-detected blazars \citep{sobolewska14}, allowing also to estimate characteristic variability time scales in addition to PSD slopes. A good description of different time domain analysis methods used in $\gamma$-ray astronomy can be found in \cite{rieger19}.

\subsection{VLBI}\label{vlbi}
Currently, the only wavelength range where blazars can be spatially resolved is in the radio and millimeter waves through interferometry. While connected-element interferometers, such as the Jansky Very Large Array (JVLA) and Atacama Large Millimeter Array (ALMA) are sufficient for studying the kpc-scale jets in radio galaxies in arcsecond-scale angular resolution (covered in Chapter 3), in order to spatially resolve the emission in the blazar zone on parsec scales, corresponding to an angular resolution of milliarcseconds, very long baseline interferometry (VLBI) is required. In VLBI observations, radio telescopes around the world are simultaneously observing the same targets, forming an interferometer with an angular resolution determined by the distance between the telescopes and the observing frequency. Typically, angular resolutions ranging from some tens of microarcsecond to some milliarcseconds can be achieved. The observations are time stamped and usually recorded on disks, after which the signal from the different telescopes can later be combined in a correlator to transform the data into a format that can be used to form an image of the source.
 
Nowadays it is also possible to conduct real-time VLBI or e-VLBI where the data are sent to a correlator real time over the internet. Technical details of interferometry are beyond the scope of this review, but interested readers can find all necessary information in the comprehensive book by \cite{thompson17}

The first VLBI experiments were conducted in the mid 1960s \citep[see][for a review of the early days of VLBI]{kellermann88}. One of the most significant discoveries in VLBI science, the detection of apparent superluminal motion in the quasars 3C~273 and 3C~279, soon followed \citep{whitney71,cohen71}. The field advanced further when imaging techniques were developed to obtain high-resolution maps of blazars \citep{wilkinson77,readhead78}, showing blobs emanating from a stationary core at apparent superluminal speeds \citep[e.g.,][]{pearson81}. 

\subsubsection{VLBI surveys and blazar population studies}
\begin{figure}
\includegraphics[scale=0.63]{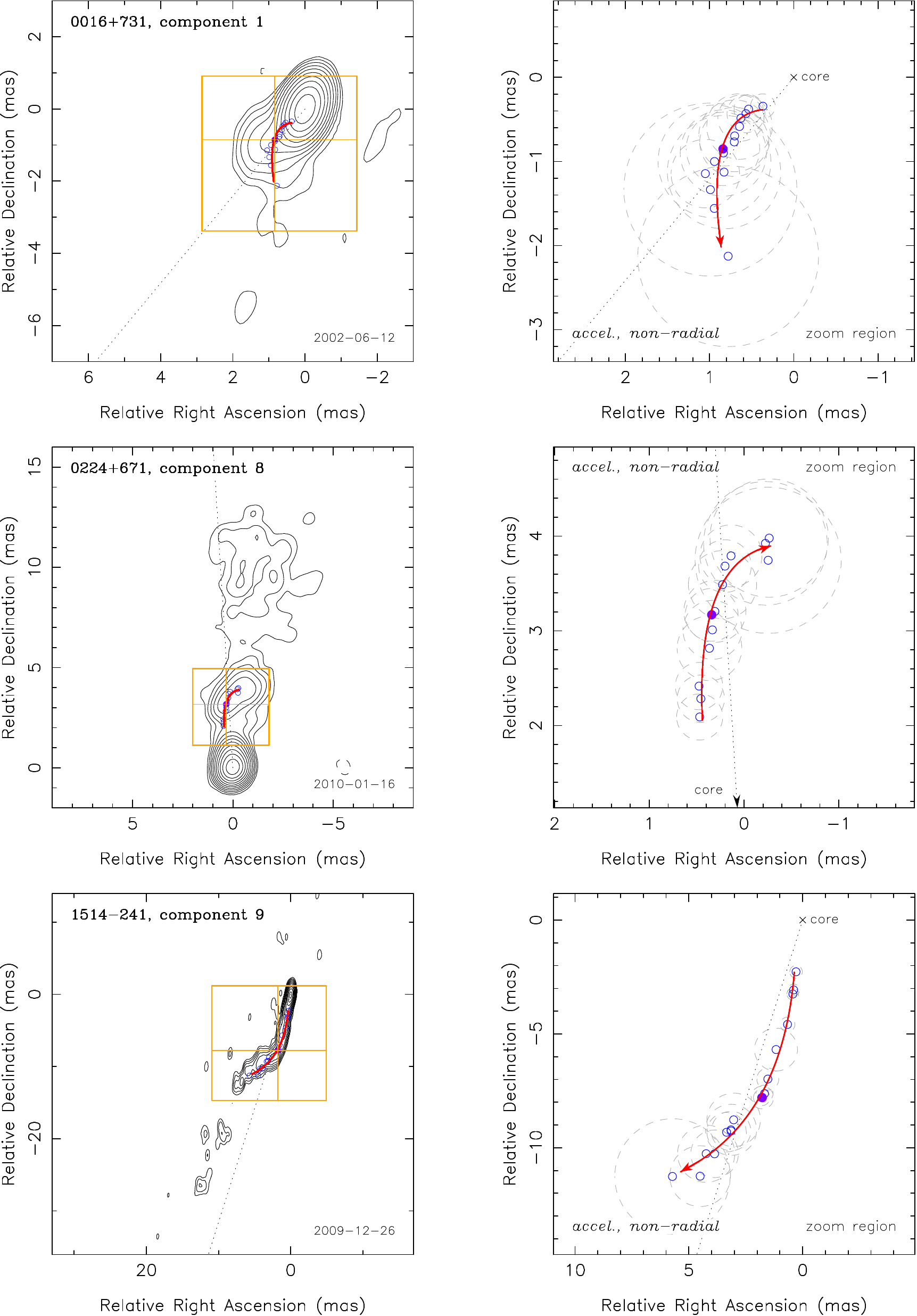}
\caption{Example images (left panels) and component motion fits (right panels) for three blazars observed by the MOJAVE program at 15\,GHz. The contours in the left-hand panels indicate the total intensity of the source on the date marked on the lower-right-hand corner of the panels. The right-hand panels show the region in the orange box. Each open circle marks the position of a specific jet component at different observing epochs, and the red curve shows the fit through these points. In all the shown cases, the components are moving non-radially from the core and show acceleration along the jet. Figure taken from \cite{lister16}. Reproduced by permission of the AAS.}
\label{fig:lister16}
\end{figure}

Another major step forward in blazar science was taken when the Very Long Baseline Array (VLBA) started operations in 1993. VLBA consists of 10 telescopes located around the USA, and it is fully dedicated to VLBI observations. This allowed the monitoring of large samples of AGN, such as the 2cm-survey program \citep{kellermann98} and its successor the Monitoring of Jets in AGN with VLBA Experiments (MOJAVE) program \citep{lister09a}, which, during the last 25 years, have imaged over 400 parsec-scale jets \citep{lister19}. At higher, 22 and 43\,GHz frequencies, obtaining a better angular resolution of down to 0.15\,mas, the Boston University (BU) blazar research group has been monitoring especially the gamma-ray loud blazars since 1993 \citep{jorstad01}. Figure~\ref{fig:lister16} left panels show example images of three blazars that have been observed by the MOJAVE program at 15\,GHz, all showing clear core-jet structures.

These surveys have shown that superluminal motion, predicted by \cite{rees67}, is very common in blazars, and it is nowadays generally accepted that the fast variability seen in blazars is a result of Doppler beaming enhancing the variations caused by changing physical conditions. This is further supported by the high observed brightness temperatures \citep[e.g.,][]{horiuchi04, kovalev05} in excess of the inverse Compton catastrophe limit of $10^{12}$ Kelvins \citep{kellermann69}. As suggested by \cite{readhead78b} and \cite{scheuer79}, relativistic beaming will result in the dominance of relativistic sources in any flux-limited samples, making blazars the most numerous sources in any surveys targeting the brightest radio sources on the sky. Beaming also makes interpretation and comparison of the observations to theoretical models challenging, because it affects the appearance of the sources.

The intrinsic properties of the blazar population can be recovered by using the observed luminosity, apparent speed, and redshift distributions \citep[e.g.,][]{urry84, urry91, vermeulen94, lister97}. Using a carefully constructed flux-density limited sample, \cite{lister19} compare the observed apparent speed, redshift and luminosity distributions from the MOJAVE sample with a population model, where the unbeamed parent population's Lorentz factor distribution follows a power-law function. The observed flux-density limited sample, which can be considered as a typical blazar sample, has a broadly peaked Lorentz factor distribution where the peak is between $\Gamma=5$ and $\Gamma=15$ with a rapid falloff at $\Gamma > 20$, indicating that using a single value for the Lorentz factor in any modeling is not observationally justified.

The objects are oriented at angles less than $\sim 10^\circ$ with a peak at $\sim 2^\circ$ from the line of sight. The unbeamed luminosities match those of powerful FR II galaxies with radio luminosities between $10^{25}-10^{26}$\,W~Hz$^{-1}$. 

Perhaps a more unexpected finding has been that many jets show acceleration of the flow even on parsec scales (see Fig.~\ref{fig:lister16} right-hand panels for examples), up to de-projected distances of 100 pc from the base of the jet \citep{homan15}. This is contradictory to most theoretical models, which assume that the jet flow has been fully accelerated within the so called {\it acceleration and collimation zone} that extends up to the radio core \citep[e.g.,][]{potter13}. Beyond these distances, the jets begin to decelerate, as would be expected based on subluminal speeds in the kpc-scale jets \citep[e.g.,][]{arshakian04}. Recently, \cite{giannios19} proposed a solution to this dilemma through a striped-jet model where the jet is magnetically launched and accelerated. The toroidal field in the jet forms stripes of different length scales related to the growth time of the magnetorotational instability in the accretion disk. This results in a gradual acceleration of the jet up to distances of 100~parsecs in blazars with ultrarelativistic jets ($\Gamma >> 1$) and black holes of $\sim10^8$M$_\odot$, which could explain the results from the VLBI observations. However, it is still unclear, without more detailed comparisons, whether this model can explain other general features of the jets, such as their magnetic field structure.

\subsubsection{Highest angular resolution observations}\label{highres}
In the past years, much higher angular resolution in the cm-wavelengths has been achieved through space interferometry where one radio telescope is placed in a satellite orbiting the Earth. One of the first such missions was the Japanese-lead VLBI Space Observatory Programme (VSOP; \citealt{hirabayashi98}), which used a 8-m telescope onboard the {\it HALCA} satellite. In addition to showing that the cores of blazars can reach brightness temperatures in excess of the inverse Compton limit \citep{tingay01, horiuchi04}, the VSOP images, for example, confirmed that some jets are clearly limb brightened, indicating the spine-sheath structure of the jet \citep{giroletti04}. 

The highest angular resolution in the cm-wavelengths has so far been reached with the Russian RadioAstron Space Radio Telescope, which was launched in 2011 \citep{kardashev13}. In addition to confirming the extreme brightness temperatures \citep[e.g.,][]{kovalev16}, it has also showed superb details of the jets, especially in nearby radio galaxies. Observations of the nucleus of the radio galaxy 3C~84 by \cite{giovannini18}, show a limb-brightened structure reaching all the way down to 30\,microarcseconds ($\mu$as) from the core. More interestingly, the width of the limb-brightened structure indicates that either the jet expands much faster than theoretically predicted for a black-hole powered jet, or the jet is launched from the accretion disk. This is in contrast with observations of the nearby radio galaxy M87 with a black-hole powered jet \citep{nakamura18}, which shows that with the highest angular resolutions, the emerging picture and unification of radio galaxies is more complex than simple models predict. 

RadioAstron has also allowed the imaging of blazars, and \cite{gomez16} show the total intensity and polarization structure of BL~Lac at the angular resolution of $21\mu$as, the highest achieved to date. Interestingly, their 22\,GHz RadioAstron image shows indications of emission $\sim50\mu$as upstream of the radio core (i.e., closer to the black hole), which they interpret as a possible recollimation shock (see also Section~\ref{coreblobs}). Their most striking results come from polarization observations (see also Section~\ref{faraday}), which when combined with ground-based VLBI observations at 15 and 43\,GHz, show indications of a helical magnetic field threading the radio core (see Section~\ref{Bfields}).
%(see Fig.~\ref{fig:rm}). 

Another way to improve the angular resolution is by going to shorter wavelengths. The first 3\,mm (86\,GHz) VLBI observations were conducted already in 1981 \citep{readhead83}, showing that the emission in the radio galaxy 3C~84 is more compact at these scales than at longer wavelengths. Since then, observations with the Coordinated Millimeter VLBI Array \citep{rogers95} and its successor the Global Millimeter VLBI Array \citep{krichbaum06} have achieved an angular resolution of some tens of microarcseconds. Millimeter VLBI observations were recently reviewed by \cite{boccardi17} where the reader can find an excellent description of the past and present achievements in the field. 

The next major breakthrough is expected to come from observations with the Event Horizon Telescope (EHT) \citep{eht2}, which combines a set of millimeter-band telescopes to achieve the highest angular resolution by shortening the wavelength down to 1.3\,mm. The main advantage of going to such a short wavelength is the reduced opacity in the jets. Most blazars are seen to transform into optically thin at around 100\,GHz \citep{planck11}, meaning that at a frequency of 230\,GHz (1.3\,mm) it is possible to view the innermost regions from where emission originates. The main scientific goal of the EHT is thus to image the shadow of the black holes in the center of Milky Way, Sgr A*, and in the nearby radio galaxy M87 \citep{eht1}.

The first results from observations in 2017 when ALMA was included in the array were recently published by the Event Horizon Telescope Collaboration \citep{eht1,eht2,eht3,eht4,eht5,eht6}. These were the first observations where it was possible to obtain an image of the black hole shadow \citep{eht3}, while in earlier observations \citep[e.g.,][]{doeleman08, doeleman12,johnson15, kim18,lu18} only constraints on the size of the event horizon were obtained from fitting the amplitudes obtained between different antennas (visibility amplitudes in VLBI jargon). The amazing results for M87 show that images with extremely good angular resolution of 20$\mu$as can be achieved.

In addition to M87, also the blazar 3C~279 was observed as a calibrator, and first images of it were shown in \cite{eht4}. The detailed analysis of the first images of 3C~279 with $\sim20\mu$as angular resolution by \cite{kim20} show two components separated by about 100$\mu$as, and while the second component is in the direction of the jet seen at lower frequencies \citep[e.g.,][]{lister16,jorstad17}, the first "core" component is perpendicular to it. The core component shows elongated structure that can be modelled with three bright features separated by $\sim30-40\mu$as. The morphology can be interpreted as a broad resolved jet base or a spatially bent jet. \cite{kim20} also report that they detect day-to-day structural changes in the jet, which are consistent with apparent speeds of $\sim15-20c$. These speeds are similar to those seen at lower frequencies \citep[e.g.,][]{lister16,jorstad17}, indicating that in this source, no acceleration is taking place between the emission sites at cm- and mm-wavelengths. We expect that many more exciting results will follow the first results published so far (see also Sect. \ref{outlook}).

\subsection{Multi-messenger observations}\label{multimessenger}

Cosmic rays were discovered just slightly before astrophysical jets, and in 2012 hundred years since the discovery of cosmic rays was celebrated. As cosmic rays are charged particles, they loose their track in the galactic and intergalactic magnetic fields and therefore arrive to earth uniformly from all directions. As it is impossible to trace their origin, the mystery of the origin of cosmic rays has persisted throughout the decades \citep[see][for a recent review]{dawson17}. In energies exceeding $10^{18}$ eV, jets of active galactic nuclei have long been one of the main candidates \citep[][]{biermann87}. While the most recent work by Pierre Auger observatory shows that cosmic rays at those energies are most certainly of extragalactic origin \citep{aab17}, pinpointing the sources to astrophysical jets launched by supermassive black holes in the center of galaxies has been very challenging.

Astrophysical ultra-high-energy neutrinos were discovered in 2013 by the IceCube neutrino observatory \citep{aartsen13}, and a bit less than one hundred neutrinos likely to be of astrophysical origin have been detected ever since \citep{icrc17}. The origin of these neutrinos is still unknown, but recently IceCube detected the high-energy neutrino IC-170922A with good angular resolution, spatially and temporally coincident  with the flaring blazar TXS~0506+056 \citep{aartsen18,ansoldi18}. An association of ultra-high-energy neutrinos to blazars had been suggested before \citep{kadler16,lucarelli17} with marginal significance (due to insufficient angular resolution of the neutrinos and the absence of $\gamma$-ray signals well correlated in time). The detection of astrophysical neutrinos from a blazar also means that protons must be accelerated to high energies in the blazar jets. Neutrinos are the end products of a proton-photon interaction, which requires energetic protons to be present, and the discovery has triggered significant interest in hadronic emission models (see Section~\ref{hadronic}). 

Finally, blazars are interesting sources for gravitational wave astronomy, as some of them are thought to host supermassive black hole binaries with a rather small separation \citep{sillanpaa88}. The gravitational waves produced by supermassive black hole binaries are in the sub-microHertz regime and not in the bands detectable by ground-based laser interferometers. However, the signal could be detectable with the space-born laser interferometer LISA (see Section~\ref{lisa}). 

\section{Open questions}\label{questions}
In this section, we list some of the most relevant open questions still remaining in blazar science, despite more than 50 years since their discovery. We also include discussion on how observations have been used to constrain the models, and what more needs to be achieved to fully answer the questions.

\subsection{Is there a blazar sequence?}

Blazar sequence was first suggested by \citet{fossati98}, who investigated a set of spectral energy distributions from radio to $\gamma$-rays (using EGRET data). They found a systematic trend in the SEDs as a function of radio luminosity, namely the highest radio luminosity objects had the lowest SED peak frequencies. The existence of the sequence has been disputed ever since, as there have been contradicting sources discovered (sources with high luminosity and high peak frequency \citealt{padovani03,padovani12}), and claims that the sequence is just an observational bias \citep{nieppola08,giommi12b}. 
\citet{meyer11} suggested that instead of a sequence, blazars form an envelope, in which the different blazar subclasses are formed due to the progressive misalignment of two intrinsically different populations of blazars (FRI and FRII radio galaxies). However, high luminosity, high peak frequency sources, such as PKS~1424+240, do not fit in the envelope scenario either \citep{cerruti17}. Also, sources with low synchrotron peak frequency and a very broad second peak of the spectral energy distribution \citep[e.g.][]{aplib} do not follow the general trends expected from the sequence and envelope scenarios.

\begin{figure}
\includegraphics[width=\hsize]{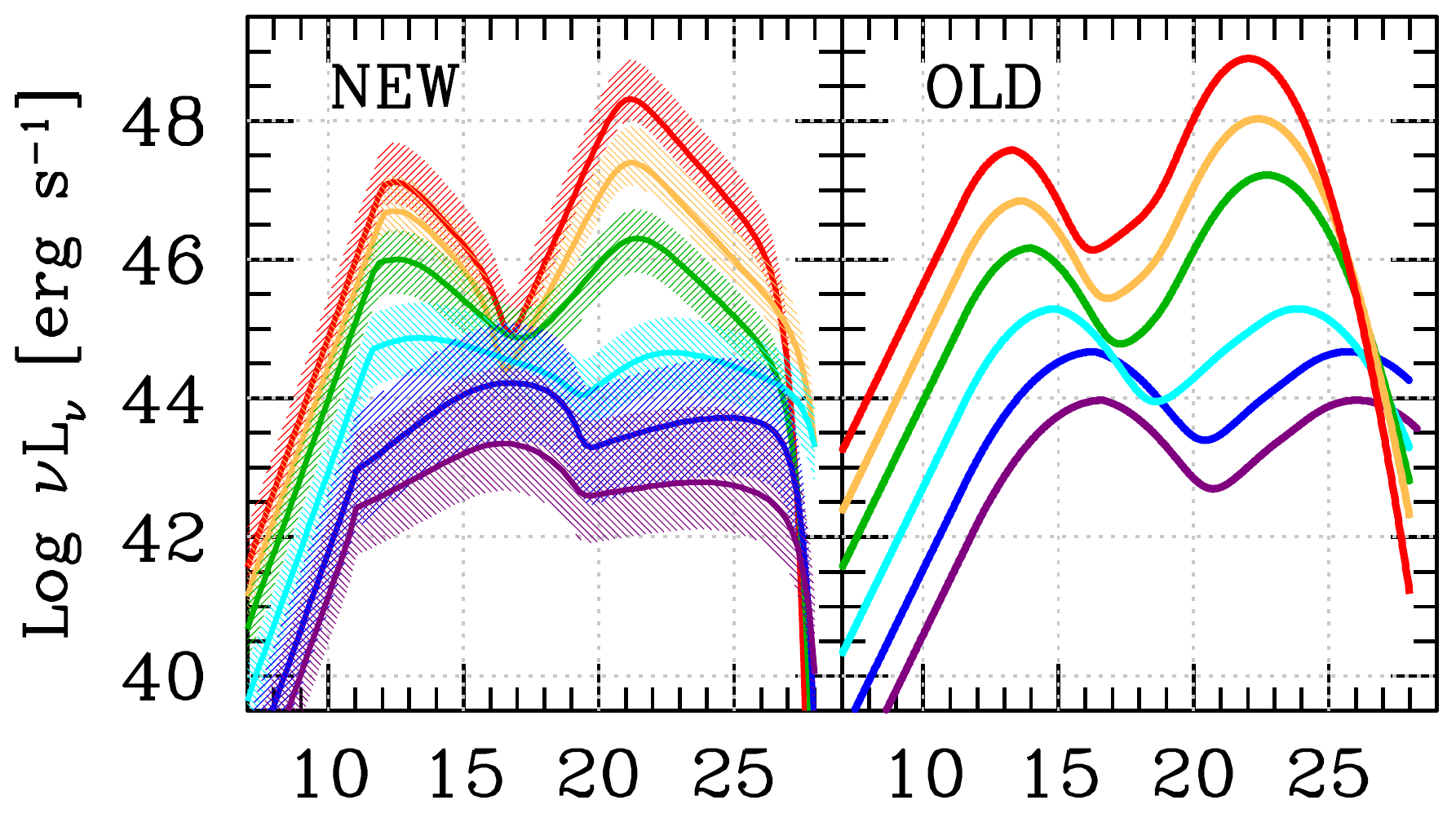}
\caption{The new \citep{ghisellini17} and the old \citep{fossati98} blazar sequence. Figure taken from \cite{ghisellini17}.}
\label{fig:ghisellini17}
\end{figure}

\citet{ghisellini17} found that the sequence indeed exists for BL~Lac objects, while for FSRQs the peak positions in the SED remain constant in different luminosity bins.
In BL~Lacs the two SED peaks have almost equal luminosities, while in FSRQs the SED is typically Compton dominated, i.e. the second peak is significantly higher, see Figure~\ref{fig:ghisellini17}. In the  "FSRQ sequence" only the amount of Compton dominance changes \citep{ghisellini17}, see Figure~\ref{fig:ghisellini17}. 
The difference between the FSRQs and BL~Lacs is interpreted to be the difference in the dominance of radiative cooling, as FSRQs have a broad line region and an infrared torus that are very efficient in cooling if the emission region is located inside these structures. The higher synchrotron peak observed in some FSRQs during the flares is interpreted as the main emission region moving outside the broad line region and dusty torus and therefore cooling becoming inefficient \citep{ghisellini13}.

The question of a sequence or an envelope closely relates to the classification of blazars as BL~Lacs or FSRQs. The original classification scenarios used the equivalent width of emission lines, EW$<5$\AA as the dividing limit \citep{stickel91,stocke91}. BL~Lacs were thought to have weak emission lines due to the very bright jet continuum washing out the emission lines \citep{blandford78}, but then many X-ray bright BL~Lacs (which nowadays are called HSPs), were found to show host galaxy features, but still very weak lines. This would indicate that the lines are intrinsically very weak. According to \citet{giommi13} these two possibilities are not exclusive, and there exists sources that have been classified as BL~Lacs due to the heavy dilution of the broad lines, but that are actually the missing high-synchrotron peaked quasars.

\begin{figure}
\includegraphics[angle=-90,width=\hsize]{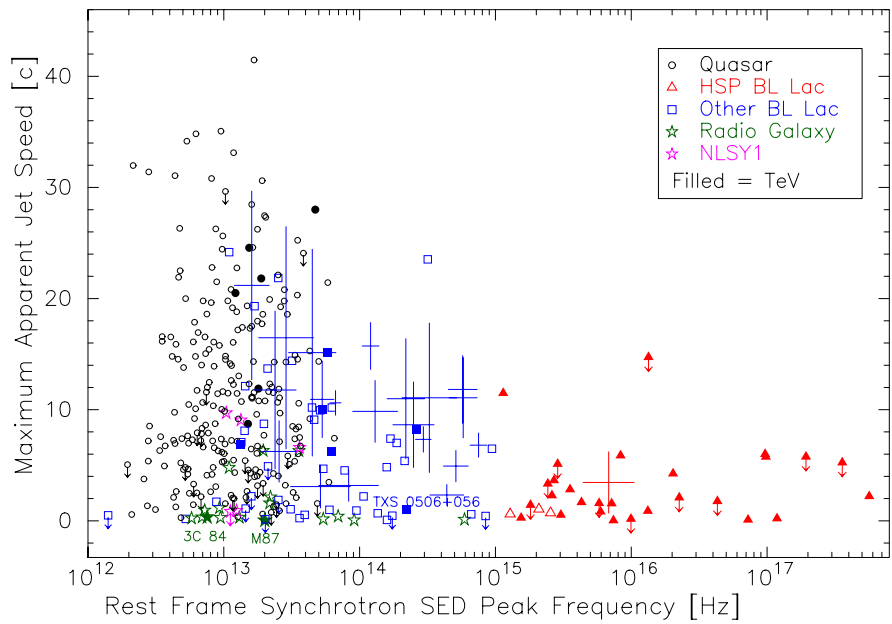}
\caption{Maximum apparent jet speed from MOJAVE 15\,GHz observations against the synchrotron peak frequency, with the different classes of sources shown with different symbols. The cross symbols indicate BL~Lacs for which only upper and lower limits on the redshift are known. Noteworthy is the much lower speeds seen in HSP sources. Figure taken from \cite{lister19}. Reproduced by permission of the AAS.}
\label{fig:lister19}
\end{figure}

The sequence and classification of the sources have also been studied by comparing the radio and $\gamma$-ray brightness of blazars \citep{lister11}. They found that BL~Lac objects display a linear correlation of increasing $\gamma$-ray loudness with synchrotron SED peak frequency, suggesting a universal SED shape for objects of this class. They also found that the high-synchrotron peaked (HSP) BL~Lac objects are distinguished by lower than average radio core brightness temperatures, and none display large amplitude radio variations or high linear core polarization levels. They found no such trends for FSRQs. They concluded that HSP BL~Lac objects have generally lower Doppler factors than the lower-synchrotron peaked BL~Lac objects or FSRQs.

The differences between FSRQs and HSP sources is clear also in radio kinematics \citep[][see Fig.~\ref{fig:lister19}]{lister19}. An independent analysis of the jet kinematics from the data of the MOJAVE program was performed by \citet{hervet16}, who classified 161 sources based on radio kinematics only, and found that the kinematic classification agrees very well with the usual spectral classification. They found that {\it class I}, with quasi-stationary knots, corresponds to HSPs and {\it class II} with knots in relativistic motion from the radio core, to FSRQs. The intermediate sources showing both {\it class I/II} features, mainly belong to traditional ISP/LSP classes. 

In summary, all the classification methods seem to agree that there seem to be clear FSRQs sources and clear HSP sources with intrinsically different nature, but the role and nature of the intermediate and outlier sources is still very unclear and under debate. Furthermore, we might learn something new about the existence of a sequence and the underlying physics, when we will extend it to even higher energies. Currently the understanding of the VHE $\gamma$-ray blazar population is still incomplete. For example, the extremely high synchrotron peaked sources have the second peak beyond the {\it Fermi}-LAT band, which is challenging to explain with traditional emission models. The first mini-catalogue of these objects was recently published \citep{acciari2020} and it consists of ten such sources. Further observations are certainly needed to understand how numerous these objects are, and if they provide additional challenges to blazar sequence models.

\subsection{Blazar zone}
This subsection combines open questions on the nature and location of the {\it blazar zone}, i.e. the region(s) where most of the emission originates, and how it can be observationally constrained.

\subsubsection{How and where is the jet converted from magnetically to particle dominated?}\label{magnetization}
Despite decades of blazar studies, one major dilemma is the connection between jets that, according to theoretical models, are launched as magnetically dominated \citep[e.g.,][]{blandford77}, where the bulk of the energy is contained in the magnetic fields, but where the emission can be best modeled by equipartition between the magnetic and particle energy densities in the jets \citep[e.g.,][]{readhead94}. Often, it is assumed that the jets start off magnetically dominated with a parabolic shape and reach equipartition at the point where the flow has been (magnetically) accelerated to terminal velocity, after which the shape of the jet is conical \citep[e.g.,][]{ghisellini89,potter13}. This is also supported by magnetohydrodynamic jet launching simulations of Blandford-Znajek-type jets \citep{mckinney06}. The question is also closely tied in with the particle acceleration mechanism in the jets (see the specific chapter in this volume). In magnetized jets, the working mechanism could be magnetic reconnection \citep[e.g.,][]{giannios06}, while in particle dominated jets, shock acceleration is expected \citep[e.g.,][]{kirk87}. 

Traditionally, the variability in radio to optical bands has been successfully modeled by shocks moving down the jets \citep[e.g.,][]{marscher85,hughes85}, which require particle dominance in the jets. This means that if the jets are magnetically launched, they must transform to particle dominated by the regions where most of the optical and radio emission originates. Observations of the jet shape on parsec scales also support the idea that most of the radio emission originates in conical jets \citep[e.g.,][]{pushkarev17}, although in some nearby radio galaxies that can be spatially resolved down to small scales, the jets appear parabolic up to a transition point, which in M87 corresponds to a distance of $10^5$ Schwarzschild radii \citep{asada12}.

Most of the early works on jet emission concentrated on conical jets, following \cite{blandford79}, who showed that a conical jet with shocks produces a flat radio spectrum. When $\gamma$-ray data became more ample in the EGRET and especially in the {\it Fermi} era, the jet modeling has shifted more to time-dependent models of individual emitting blobs (so called single-zone models, see Section~\ref{sedparameters}). A full model for a quiescent jet is presented by \cite{potter13} who model the jet as initially magnetically dominated, transitioning to a conical jet in equipartition. While their model can adequately fit the SEDs of sources in quiescence, including their radio emission that is often ignored in single-zone models \citep{potter13b}, it does not account for flaring, which is typically the initial starting point for the single-zone models. Their model requires the transition to occur at fairly large ($>10$ pc) distances from the black hole, corresponding to $10^5$ Schwarzschild radii in M87, which means that the transition occurs further out than the edge of the dusty torus \citep[e.g.,][]{nalewajko12}. This would indicate, that in SED models requiring external photon fields to explain the Compton peak, the photon field must be far out (as is expected by \citealt{potter13}), or the flaring occurs in the magnetically dominated region, which contradicts with the typical assumptions of SED models. 

In the recent striped-jet model by \cite{giannios19}, the jet is magnetically accelerated up to the point where it reaches its maximum Lorentz factor. For a typical black hole mass of $\sim10^8$M$_\odot$ this occurs at a distance of about 5\,pc from the black hole. Interestingly, their model indicates that dissipation in the jet can occur over a broad range of distances ($\sim 0.03-300$\,pc for an ultrarelativistic jet reaching $\Gamma_\mathrm{max}\sim30$), with shorter variability time scales expected at distances closer to the black hole. While their model can explain many of the general features of blazar jets (flat spectrum, continuous acceleration, different time scales), there have not yet been any detailed fits to data of individual objects.

Thus, it is still not clear how the jets are accelerated and at what point they transform from magnetically dominated to equipartition. Because the magnetohydrodynamic jet launching simulations do not, by definition, include any radiating particles, it is impossible to make direct comparisons between the simulations and observed radiation. The simulations also typically assume self-similar jets, which is clearly an invalid assumption considering the flaring behavior observed. Therefore, some new developments combining the large-scale fluid physics with the microphysics of particle acceleration, in addition to high-quality multifrequency observations, are required to eventually solve this problem. 

\subsubsection{What is the nature of the radio core and blobs in jets?}\label{coreblobs}
As discussed in Section \ref{vlbi}, the jets of blazars exhibit components or blobs that move down the jet at apparent superluminal velocities. They are seen to emanate from the bright, typically unresolved, end of the jets called the core. The nature of this core is not always clear. It has been suggested that at higher frequencies corresponding to mm-wavelengths, the core would be a standing shock \citep[e.g,][]{daly88}, which are seen to naturally form in hydrodynamic \citep[e.g.,][]{gomez95,mimica09} and magnetohydrodynamic \citep[e.g.,][]{mizuno15,barniol17} jet simulations. At lower cm-band frequencies, the core is often seen to correspond to the surface where the jet becomes optically thin at that frequency \citep{marcaide84}, i.e. its location is not fixed but varies as a function of frequency. This corresponds to the traditional Blandford-K\"onigl type jet \citep{blandford79}, which is a conical synchrotron self-absorbed jet, where the magnetic energy density and particle energy density are in equipartition. 

The nature of the core can be studied by observing the sources at multiple frequencies, and carefully aligning the images so that the absolute position of the core, and its possible frequency dependence, can be examined \citep{marcaide84,lobanov98}. In the cm-wavelengths, this has been done for a large number of sources \citep[e.g.,][]{kovalev08,sokolovsky11,pushkarev12}, and the location of the core in most sources seems to follow the relation $\nu^{-1}$ as expected for a $\tau=1$ surface in a Blandford-K\"onigl type jet \citep{blandford79}. The core shifts can also be used to determine the distance of the core from the central black hole. \citet{pushkarev12} used the core shift measurements at 15\,GHz to calculate the distance of the VLBA core from the jet apex. The  distributions were different for  quasars  and  BL  Lacs, with medians of 13.2 and 4.0 pc, respectively.

There have been attempts to extend the core-shift studies to mm-wavelengths, where the effect is expected to disappear if the core is instead a standing shock. However, due to the difficulty in obtaining simultaneous high-quality VLBI data up to mm-wavelengths, the results are still somewhat inconclusive. \cite{marscher08} interpret the core of BL~Lac at 43\,GHz to correspond to a standing shock in the flow, because some polarized features are detected upstream (i.e. closer to the black hole) of the brightest feature called the core. Similar upstream emission was also detected in the highest angular resolution observations of BL~Lac at 22\,GHz by RadioAstron \citep{gomez16}. In the BL~Lac object 1803+784, the radial pattern of the polarization vectors in the core component is similar to what is seen in simulations of conical shocks \citep{cawthorne13}, supporting the view that the core is a standing shock.

On the other hand, in \cite{fromm15} the location of the core in the FSRQ CTA~102 was studied at $5-86$\,GHz frequencies, and they found a significant shift between the 43\,GHz and 86\,GHz core positions, indicating that even at 3\,mm wavelength, the core would still correspond to a $\tau=1$ surface.  This could indicate that there are differences in the flows of BL~Lacs compared to FSRQs. Thus, the nature of the core at different wavelengths still remains unconfirmed.

The nature of the superluminal components is also not clear. Following the discovery of variable radio emission in blazars by \cite{dent65}, a model of expanding blobs was suggested by \cite{paulinytoth66} and \cite{vanderlaan66}, based on a model by Shklovsky for expanding supernova remnants. In this model, the blobs of plasma cool adiabatically, which results in changes in the  synchrotron spectrum so that the peak of the spectrum moves to lower frequencies while the flux density decreases as the blob expands. Despite the first multifrequency observations agreeing with this simple model \citep{dent68}, it was soon noted that with more data, the simple model could no longer explain all the spectral variations \citep{altschuler77}. It was then suggested by \cite{blandford79} that the moving knots are shocks in the jet, which was supported by radio and optical observations of blazar flares \citep{marscher85} and especially by the linear polarization variability in the cm-wavelengths \citep{hughes85}.

Although the shock model and its updated versions \citep[e.g.,][]{turler00,fromm11} continue to work well for the optical and radio variations of blazars, at the highest energies, especially in VHE $\gamma$-rays the variations are often too fast to be explained by shocks. As a solution, \cite{giannios09} developed a {\it jet-in-jet} model where the variations are due to small mini-jets inside the jets of blazars. The mini-jet has a faster bulk speed of the plasma than the ambient jet plasma, and the energy dissipation is happening through magnetic reconnection. In this case, the emission regions could eventually form larger plasmoids \citep{giannios13} that could be seen as blobs of plasma in the jets. This indicates that most likely, the jets exhibit both more general plasma blobs and shock fronts where particles are accelerated.

In some cases, the jets also exhibit quasi-stationary components, which do not seem to be moving with respect to the core location \citep[e.g.,][]{jorstad05,marscher08,lister13,cohen14a,gomez16}. These are, in general, interpreted as evidence of recollimation shocks forming further down the jet \citep[e.g.,][]{gomez95,mizuno15}, similar to what the HST-1 feature in M87 could be \citep[e.g.,][]{cheung07}.

\subsubsection{What is the location of the blazar zone and the source of the seed photons for high-energy emission?}\label{seedphotons}

As discussed in the earlier sections, the only wavelength regime at which we can spatially resolve the jet is radio, where the jet becomes optically thin several parsecs away from the black hole. Therefore, the location of the blazar zone has been a topic of decades long debate. The question closely connects to the origin of the seed photons for the inverse Compton scattering, as the availability of the seed photons depends on the distance from the central engine. At least in FSRQs the central engine is surrounded by BLR clouds that re-scatter emission from the accretion disk. Further out a dusty torus surrounds the BLR.

\begin{figure}
\includegraphics[scale=0.12]{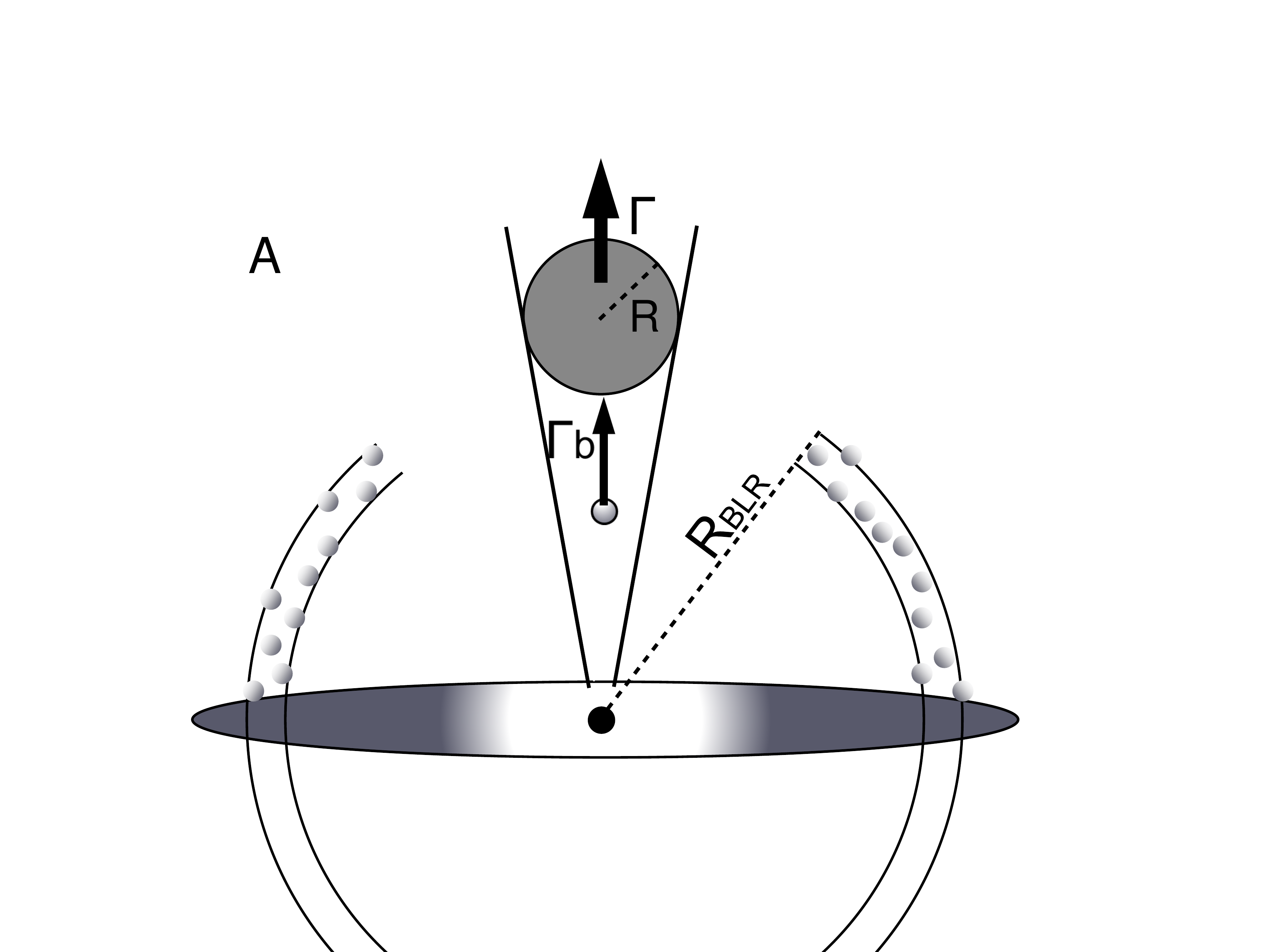}\includegraphics[scale=0.12]{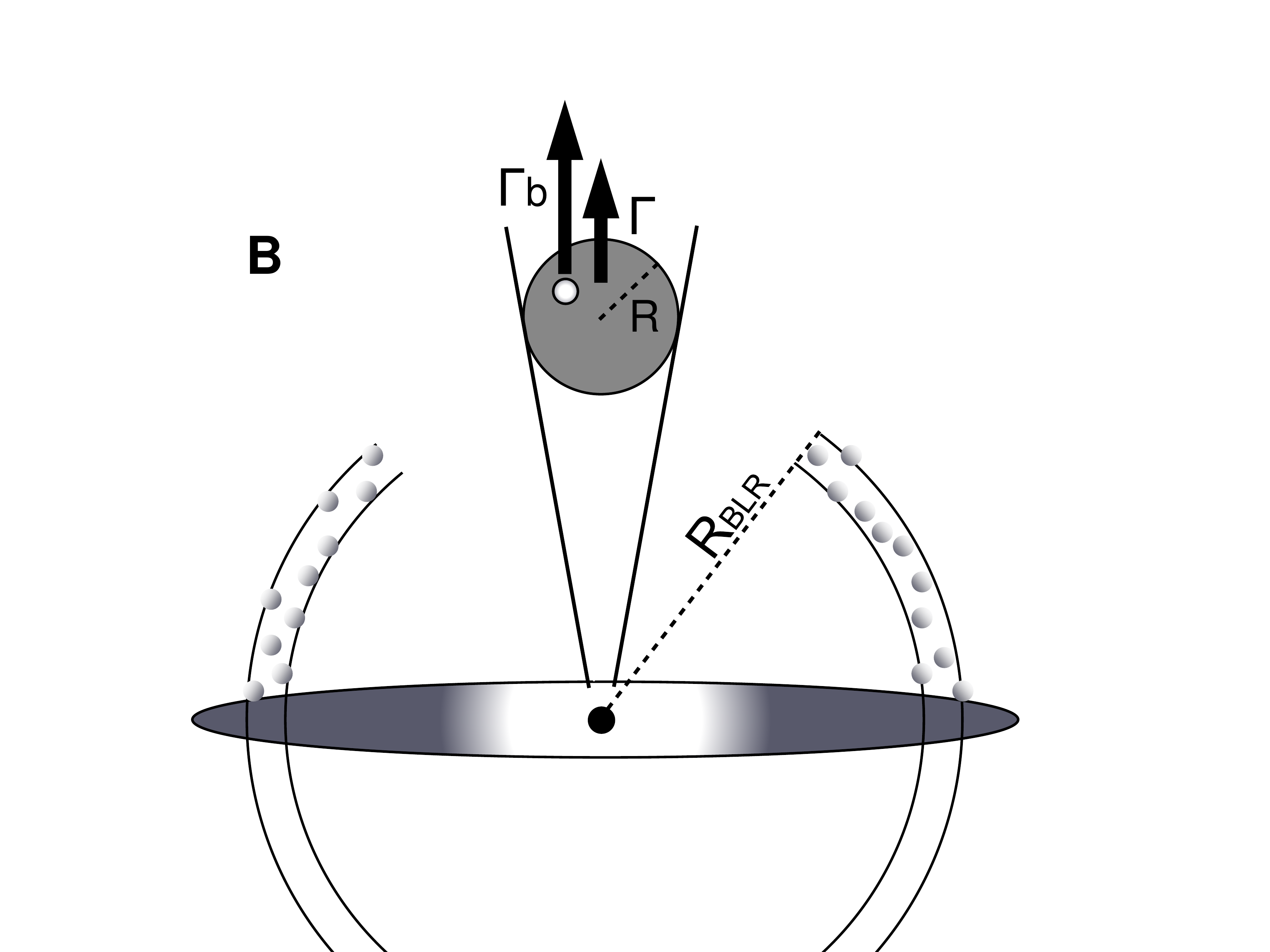}\includegraphics[scale=0.12]{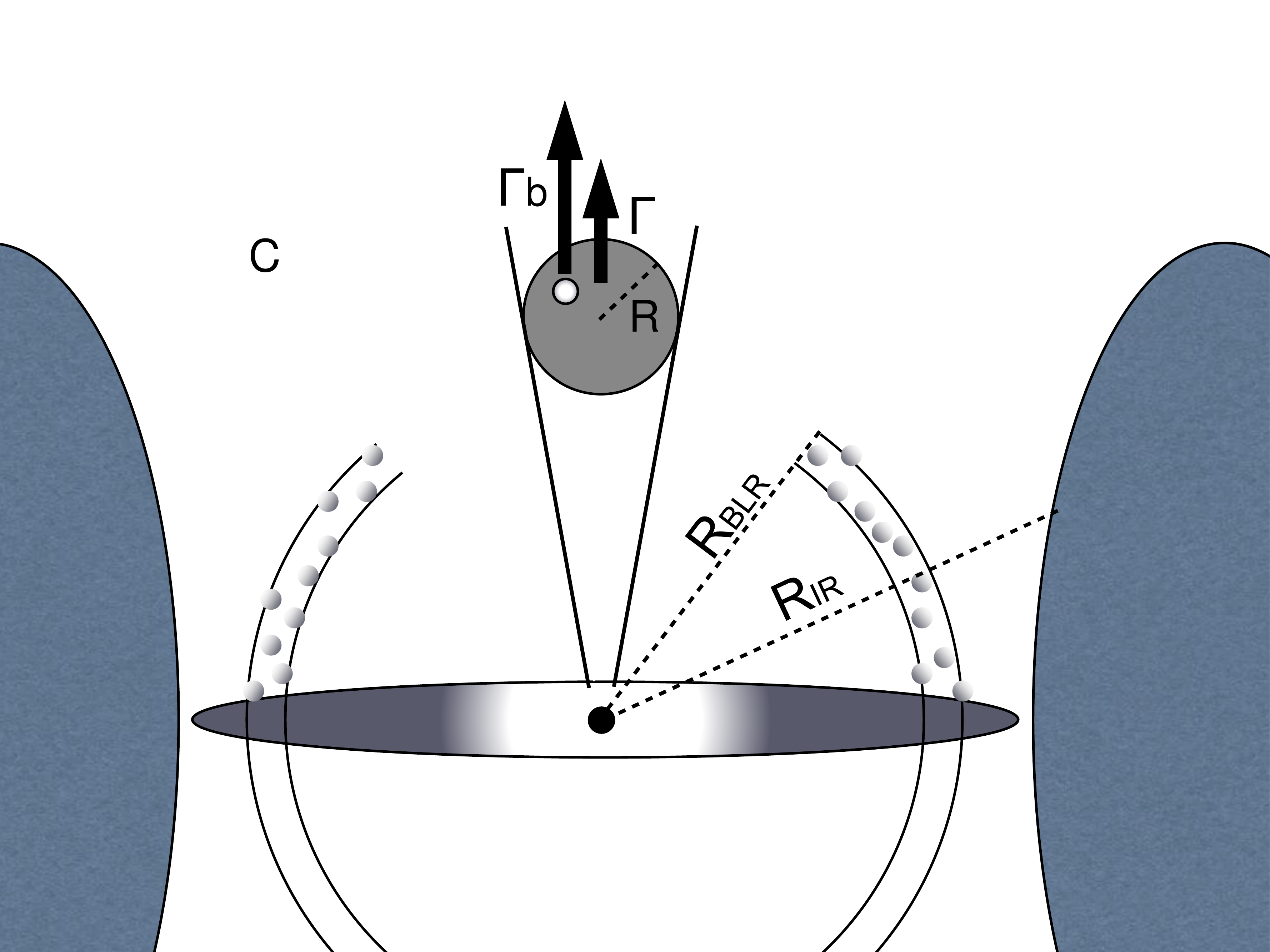}
\caption{Cartoons of the different possible locations of blazar zones, assuming two zones where the smaller emission region is responsible for fast variability in $\gamma$-rays and the larger one for the overall emission. In cartoon A the two regions are not co-spatial and the small emission region is located close to central engine, inside the BLR. In cartoon B the two emission regions are co-spatial and outside the BLR. This cartoon represents the situation for BL~Lac, where there is no observational evidence for a dusty torus. Cartoon C is as cartoon B, but for FSRQs where a dusty torus is present and can provide seed photons for inverse Compton scattering.
Figures A and B taken from \cite{acciari19}, C from \cite{tavecchio11}. Reproduced with permission from Astronomy \& Astrophysics, \copyright ESO.}
\label{fig:cartoon}
\end{figure}

During flaring states, the SEDs of FSRQs are highly Compton dominated, and therefore there must be a source of external seed photons present in the location where the flares are produced. The BLR has long been the main candidate for external seed photons for inverse Compton scattering in quasars \citep{sikora94,blandford95,dermer97,hartman01}. If the emission region is within the BLR, the photon density from the BLR is very high, and the inverse Compton scattering of those photons dominates the $\gamma$-ray emission. However, the photon density decreases very fast outside the BLR \citep[e.g.][]{bottcher16}. Therefore, the detailed geometry and size of the BLR are very relevant for emission models. They are still under debate, but values typically adopted for the extensions are of order $0.1-1$ parsec. These estimates are based on scaling between the radius of the BLR and AGN luminosity. The relation has been established using reverberation-mapping, i.e. by looking at the delay between the brightening of the emission lines and the continuum \citep{blandford82}. The scaling is widely used and carefully studied, and in the past years the slope has been well determined and the scatter reduced \citep[][and references therein]{bentz13}. Very recently, the first direct detection of the BLR was achieved by using a near-infrared interferometer at the Very Large Telescopes, GRAVITY, \citep{sturm18}, and the results were well in agreement with the results from reverberation mapping. So, in order for the emission region to "benefit" from BLR photons, it should be located within $\sim$1 parsec from the central engine (see Figure~\ref{fig:cartoon}, panel A). 

Further argument to favor the location of the blazar zone close to the central engine has been the fast $<$1 hour time-scale variability observed in $\gamma$-rays \citep[e.g.][]{saito13,foschini13}, in 3C~279 the variability time scale was as short as 5 minutes \citep{ackermann16}. The fast variability limits the size of the emission region to be small, which would place the emission region close to the black hole {\it if} we assume that the emission region fills the full volume of the jet diameter (see below for arguments overcoming this assumption).

VHE $\gamma$-rays provide a strong tool to locate the emission region. Producing VHE $\gamma$-rays requires both a dense field of target photons for inverse Compton scattering, and yet they should not be produced inside a too dense photon field in order to not get absorbed. If the blazar zone is located inside the BLR, the VHE $\gamma$-rays would get absorbed \citep{donea03,tavecchio09}. Yet, VHE $\gamma$-rays have been detected from several FSRQs, starting from 3C~279 \citep{albert08}. The detections have been made mostly during flares in other bands, but PKS~1510$-$089 also has been detected during the low activity state \citep{acciari18}. In all cases, the detection of VHE $\gamma$-rays, even at low level, locates the emission region outside the BLR. We also see fast variability from FSRQs, with a variability time scale of $\sim10$ minutes, in the VHE $\gamma$-ray band \citep{aleksic11,zacharias17}. As this indicates that the emission region must be both small and outside the BLR, it provides direct observational evidence that there must be substructures within the jet. There must be small emission regions that under some conditions can dominate the whole emission. Therefore, a small emission region does not automatically mean that the emission region must be close to the central engine.

With {\it Fermi} data, the {\it far from the black hole} scenario has received support \citep{marscher12,jorstad13}. Using VLBI observations combined with {\it Fermi}-LAT $\gamma$-ray light curves they found that the timing of the $\gamma$-ray flares is close in time with activity seen in the 43\,GHz VLBI core, indicating co-spatiality. This is in agreement with earlier results from the EGRET era \citep{lahteenmaki03,lindfors06} where timing of the bright $\gamma$-ray flares were compared with flares in the radio band. As the VLBI core at 43\,GHz is located far beyond the canonical radius of BLR, for example in PKS~1510-089 the radius of the BLR is $\sim$0.1\,pc while the 43\,GHz VLBI core is located at 6.4\,pc (see discussion in \citealt{aleksic14}), the BLR seems to be excluded as the source of the seed photons for Compton scattering. This is further supported by the non-observation of BLR absorption in the large sample of {\it Fermi}-LAT FSRQs \citep{costamante18} (see also Figure~\ref{fig:cartoon} panel B). However, there are also observations that suggests this to be too simplistic conclusion. \citet{leontavares13} observed a statistically significant increase and decrease of the MgII emission line flux coincident with superluminal jet component traversing through the radio core in 3C~454.3. Similar behaviour has been seen also in other sources \citep[most recently][]{chavushyan20}.
This crucially suggests that at least occasionally there would be broad-line region clouds surrounding the radio core, which could serve as a source of seed photons for inverse Compton scattering.

In addition to VLBI and VHE $\gamma$-ray observations, correlated variability can be used to locate the emission region within the jet. Correlation between the GeV $\gamma$-ray and optical bands is often observed in FSRQs \citep[e.g.][]{hayashida12}, while correlated flares in X-rays and $\gamma$-rays are more rare.  Simultaneous flares in two bands, however, does not reveal the location of the emission region, only that the emission in these bands originates from the same region. Finally, polarization swings, i.e. rotation of the electric vector position angle (EVPA, see also Section~\ref{Bfields}) has been used to locate the emission region \citep{marscher08, larionov08, marscher10, aleksic14}. In these cases, the swing is simultaneous with the variability of the polarization of the 43\,GHz VLBA core, and, therefore, the core is the likely site of the rotation of the EVPA. The timing of flares and rotation swings with respect to the total-intensity brightening of the VLBI core has also been used to locate the emission regions. It has been found that there are indeed multiple sites, both downstream and upstream of the VLBI core, where $\gamma$-ray flares take place \citep{rani18}. As discussed in Section~\ref{coreblobs} the nature of the 43\,GHz core is still uncertain. It could be a standing shock, but also simply a $\tau=1$ surface where the jet becomes optically thin to 43\,GHz radiation.

If the main blazar zone in FSRQs is indeed close to the 43\,GHz VLBI core, which is located some parsecs away from the central black hole, it is in principle viable that the dusty torus is providing the seed photons for the Compton scattering \citep{sikora09} (see also Figure~\ref{fig:cartoon} panel C). However, in PKS~1510-089, for example, the radius of the dusty torus is estimated to be $\sim$3.2\,pc \citep{nalewajko12}, while the 43\,GHz VLBA core is located at 6.4\,pc (see above). On the other hand, the SED of PKS~1510-089 is well described from near-IR to VHE $\gamma$-rays with an external Compton model, where the main source of the seed photons is the dusty torus \citep{saito15,ahnen17}, complicating the picture. 

Finally, if the main blazar zone is, at least occasionally, located beyond the BLR and dusty torus, the seed photons for Compton scattering could originate from a slower sheath surrounding the jet \citep{ghisellini05}. There is some observational evidence, that such structure could exist also in FSRQs \citep[][see also Section~\ref{structure}]{attridge99,macdonald15,macdonald17}. \citet{aleksic14} attempted such modeling for PKS~1510-089, and indeed it results in an equally good description of the observed SED, but it requires the magnetic field strength of the emission region to be extremely low. \citet{macdonald15} presented a more detailed calculation of a blob-sheath model, where there is a local enhancement within the jet sheath (a shocked segment/ring) that is providing the seed photons for Compton scattering. In \citet{macdonald17} it was demonstrated that such a model can reproduce the time profile of the $\gamma$-ray flares, but also in this model the magnetic field strength of the blob was rather low, even if not as low as in the simplistic model presented in \citet{aleksic14}. 

One of the possible conclusions about the current understanding of the location of the blazar zone in FSRQs is that while there seems to be general consensus among observers that it is located far from the black hole, "a number of theorists have gone into denial over this result" \citep{marscher16}. However, the two views are gradually getting closer to each other.

As discussed in the earlier sections, low and intermediate synchrotron peaked BL~Lacs share some common features with FSRQs and some with HSPs. Their SEDs are not significantly Compton dominated like in FSRQs, but the two SED peaks are rather equal like in HSPs. Some LSP sources show weak emission lines, indicating that they indeed host a BLR unlike HSP sources. For example, in BL~Lac itself weak broad emission lines have been detected \citep{corbett96,capetti10}. The VLBI jets of LSP/ISP BL~Lacs show superluminal motion like FSRQs, unlike HSPs which typically show only subluminal speeds \citep[e.g.,][]{lister19}. For the location of the blazar zone, there is growing evidence that also for LSP/ISP sources it would be located close to the 43\,GHz VLBI core. It has been observed in several sources that around the time of the VHE $\gamma$-ray activity (which is often connected to activity in the optical and $\gamma$-ray regimes in these sources like in FSRQs) also the 43\,GHz VLBI core brightens and there is at least an indication that a new component is ejected from the core \citep{ahnen18a,ahnen18b,abeysekara18,acciari19}, see also Figure~\ref{fig:0716a} and \ref{fig:0716b}. Many LSP/ISP sources also have standing shock features in the jets, and there is growing observational evidence that in some cases, the $\gamma$-ray emission region is located tens of parsecs from the central engine \citep{agudo11,pushkarev19}. In this case, of course, there are no other seed photons available than the synchrotron photons themselves or a sheath surrounding the jet.

In HSP sources and the sources that are borderline ISP/HSP sources, it is very difficult to locate the emission regions. The VLBI jets typically appear subluminal \citep[][and references therein]{piner18} and there are no apparent components emerging from the VLBI core. On the other hand, these sources are typically bright VHE $\gamma$-ray sources, and to produce the observed flux and variability, high Doppler factors are required. This would indicate that the VHE $\gamma$-ray emission region is not the one we see with VLBI observations. Therefore, there are not many constraints for its location. The VHE $\gamma$-ray emission is produced co-spatially with the X-ray emission site, as most of the sources show correlation between X-rays and VHE $\gamma$-rays with small or no time lag \citep[e.g.][]{veritas_mrk421,magic_mrk421}. The optical emission, on the other hand, seems to be partially originating from the VHE $\gamma$-ray -- X-ray region and partially from the VLBI core \citep{lindfors16}. Traditionally, it has been assumed that it is located closer to the black hole than the VLBI core and the jet is decelerating \citep{georganopoulos03}. Another possibility in this case as well is that there is a spine-sheath structure \citep{ghisellini05}, where the spine is emitting the VHE $\gamma$-rays and X-rays, and the sheath is visible in the radio band.

In summary, the past ten years have shed a lot of new light on the location of the blazar zone, and it has become very evident that there are multiple zones. For FSRQs and LSP/ISP sources "the main site" is frequently close to the 43\,GHz VLBI core, but certainly not always, as we also see $\gamma$-ray flares without activity in the 43\,GHz VLBI core \citep[e.g.][]{lindfors15}. For HSPs, it is still an open question, if the two (or more?) emission components that we see in the variability and SEDs are co-spatial or not.

\subsection{SED modeling parameters from observations}\label{sedparameters}

In the simplest form, the emission region of blazars is modelled as a spherical blob filled with electrons distributed in energy according to a smoothed broken power law:
\begin{equation}
N(\gamma)=K\gamma^{-n_1}\left(1+\frac{\gamma}{\gamma_{\mathrm b}}\right)^{n_1-n_2},   \gamma_{\mathrm min}<\gamma<\gamma_{\mathrm max}.
\end{equation}
The distribution has a normalization $K$ between $\gamma_{\mathrm min}$ and $\gamma_{\mathrm max}$ and slopes
$n_1$ and $n_2$ below and above the break, $\gamma_{\mathrm b}$ \citep{maraschi03}. The emission region has a magnetic field $B$, size $R$,
and a Doppler factor $\delta$. In addition, if there is an external photon field present, its luminosity enters the calculation. The parameters have some degeneracy, for example, a larger emission region and an increasing $K$ both result in a higher luminosity of both SED peaks. 

Independent of details of the emission model itself, it is of interest to be able to constrain these parameters from observations. \citet{tavecchio98} described a simple scenario how $\gamma_{b}$, $B$ and $\delta$ can be solved from frequencies of the synchrotron and Compton peaks, $\nu_S$, $\nu_C$ and their luminosities $L_S$ and $L_C$. The problem was that, at that time, these were not always extremely well constrained from the observations. This has significantly improved with the current instruments, see Section~\ref{sed}. However, there are also other methods to constrain the jet parameters relevant to SED modeling directly from observations. We will describe these below.

{\bf Size of the emission region, R:} As discussed in the previous subsection, the variability time scale ($t_{var}$) constrains the size of the emitting region: $R\leq ct_{var}\delta(z+1)^{-1}$. Typically, the variability time scales are shortest in the $\gamma$-ray bands, and longer in lower energies, which indicates that the size of the emission region is energy dependent. However, for simple SED modeling as described above, one typically uses the shortest time scale, because it usually also dominates the energy output of the source. The size of the radio emitting region can also be estimated from VLBI observations if the VLBI core is resolved. Its size can be used to estimate the size of the emission region as in blazars the core typically dominates the emission in the radio band. Some recent papers using two-zone SED models have used the size of the core as the size of a bigger emission region, and derived the size of the smaller region from the variability time scale in X-rays or VHE $\gamma$-rays \citep{aleksic14b,acciari19}.

{\bf Doppler factor, $\delta$:} There are several ways to calculate the Doppler factor from observations, both from VLBI and single-dish radio observations. Using VLBI it is possible  to  directly  observe  the  brightness temperature of the source ($T_\mathrm{b,obs}$), which can be compared to the  intrinsic  brightness  temperature  of  the source  ($T_\mathrm{b,int}$) (often assumed to be the equipartition temperature $T_\mathrm{eq}\sim10^{11}$\,K \citep{readhead94}).
The excess in $T_\mathrm{b,obs}$ is interpreted to be  caused  by  Doppler  boosting \citep{kellermann69}. Another way to estimate the Doppler factor is to use variability. \citet{jorstad05} calculated the flux decline time ($t_\mathrm{obs}\propto t_\mathrm{int}\delta$) of a component in the jet, and compared it to the measured size of the VLBI component. Assuming then that the intrinsic variability time scale $t_\mathrm{int}$ corresponds to the light-travel time across the knot, the Doppler boosting factor can be estimated. \citet{lahteenmaki99} and \cite{hovatta09} estimated the variability time scales and variability amplitudes of the flares from the total flux density observations, which can be used to calculate Doppler factor if the  $T_\mathrm{b,int}$ is again assumed to be $T_\mathrm{eq}$. Different methods are compared in \citet{lahteenmaki99}. 

These studies have concentrated on FSRQs and radio-selected BL~Lacs (i.e. mostly LSPs). \citet{hovatta09} calculated $\delta$ for 87 sources and found an average value of 14.6 for FSRQs and 6.3 for BL~Lac objects. Recently, \citet{liodakis18} derived variability Doppler factors for 1029 sources observed by the Owens Valley Radio Observatory 40-m Telescope at 15\,GHz. They found the median $\delta \sim 11$ for blazars, but they did not find significant differences between FSRQs and BL~Lac objects, possibly due to the much larger sample of objects than in earlier studies. The sample was still dominated by FSRQs and LSP BL~Lacs, and had fewer HSPs, as they are weak in the radio bands and rarely show clear flares that could be used to estimate variability time scales. \citet{hovatta15} derived a variability Doppler factor for Mrk~421 during the so far largest detected radio outburst of the source in 2012, but even in this case they found $\delta\sim4$. In VLBI observations, HSPs show subluminal speeds or even no motion \citep[][and references therein]{piner18}, and \citet{piner18} derived an upper limit $\sim$4 for the bulk Lorentz factors. 

It is typical that the values for Doppler factors used in the SED modeling are higher than derived from radio observations, especially for HSP sources. For HSP sources this is dubbed as the {\it Doppler factor crisis} as the SEDs simply cannot be modelled with $\delta$ values as low as derived from radio observations. Several solutions to this problem has been suggested, such as: non-steady magnetized flows \citep{lyutikov10}, decelerating jets \citep{georganopoulos03}, and spine-sheath models \citep{ghisellini05}. However, even for FSRQs there is tension between the observed Doppler factors and the ones typically used to reproduce the observed spectral energy distributions, in particular in case of fast VHE $\gamma$-ray flares. The fast flares imply $\delta>50$, because otherwise the emission region would not be optically thin to $\gamma$-$\gamma$ absorption. For example, in the case of PKS~1222+216, the fast variability requires $\delta=75-80$ \citep[e.g.][]{tavecchio11,ackermann14}, while VLBI observations suggest $\delta\sim10$ for the jet \citep{jorstad17}.
 
 {\bf Magnetic field strength, B:} The magnetic field strength can be estimated from the VLBI core shift-measurements, assuming equal energy carried by the particles and the magnetic field, as done in \citet{pushkarev12}. The median value estimated for their sample of 18 BL~Lac objects is $0.4^{+0.3}_{-0.1}$G and for a sample of 84 FSRQs  $0.9^{+0.2}_{-0.1}$G. The magnetic field strengths are derived for the distance 1\,pc from the central black hole. These are some of the best estimates we have for the magnetic field strength from the observations. However, as said, the estimation assumes equipartition, and \citet{tavecchio16} showed that the spectral energy distributions of BL~Lacs cannot be reproduced if equipartition is assumed. On the other hand, for FSRQs, LSPs and ISPs parameters close to equipartition seemed to describe the SED well \citep{bottcher13}. Recently, also \citet{sobacchi19} questioned the need for sub-equipartition magnetic fields. Therefore, it is of utmost importance to study the magnetization of the jets with further observations (see also Section~\ref{magnetization}). 
 
 There is also another, more robust, way to derive magnetic field strength directly from VLBI observations, without having to assume equipartition. This is possible if one can measure the spectra, size, and Doppler factors of the individual components from VLBI observations like done in \cite{savolainen08} for 3C~273. Especially important is the observation of the turnover frequency of the spectrum, which means that this type of analysis can only be done in the inner regions of jets where the turnover is at sufficiently high frequencies to be observable. \cite{savolainen08} measure the magnetic field strength of the core to be $\sim1$G, with lower values further away from the black hole. Interestingly, their results indicate that the core is magnetically dominated. Unfortunately, these kind of direct measurements are very challenging to obtain, and have not been performed for other blazars. 
 
{\bf Low energy cutoff of the electron spectrum, $\gamma_{min}$:}
One observational way to constrain the low-energy cutoff of the electron spectrum, $\gamma_{min}$, is through circular polarization observations (see also section \ref{cp}). More specifically, if the circular polarization is due to Faraday conversion, the amount of circular polarization can be used to constrain $\gamma_{min}$ \citep[e.g.][]{beckert02}. By modeling the full-polarization spectrum (including total intensity, linear and circular polarization) of 3C~279, \cite{homan09} was able to constrain the low-energy cutoff to $5 < \gamma_{min} < 35$. The values typically used in SED modeling vary from 1 to $10^5$ \citep{tavecchio10}, which is obviously a much wider range than the one derived from observations. The high value of $\gamma_{min}$ ($\sim 10^{4}-10^{5}$) in SED models was originally suggested in \citet{katarzynski06} as a solution to reproduce the extremely hard SSC spectra of extreme BL~Lacs, but has later been used in "normal" BL~Lacs in combination with a soft spectral index above the $\gamma_{b}$ ($n_2$) to reproduce the narrow synchrotron peak and large separation between the two SED peaks \citep{aleksic12}. The feasibility of a high $\gamma_{min}$ ($10^3-10^4$) value has received some support from simulations of particle acceleration in relativistic shocks \citep{virtanen03,sironi11}. As discussed in section \ref{cp}, due to the typically low fraction of circular polarization, it is challenging to study the circular polarization spectrum, which is likely one reason why the observed constraints have not been used in many SED modeling attempts, another reason being that the SED models also tend to ignore the radio part of the SED, arguing that it originates from a region further out. A notable exception is the quiescent-jet model by \cite{potter13b}, and it would indeed be the best if modeling attempts could account for the full jet emission, covering both the quiescent and flaring parts simultaneously.

{\bf Energy density of the seed photon fields, U$_\mathrm{BLR}$ and U$_\mathrm{DT}$:}
As discussed in the previous subsections, only FSRQs show strong emission lines in their observed spectra. The main lines are Ly$\alpha$, C IV, Mg II, H$\gamma$, 
H$\beta$, H$\alpha$, C III, Fe II and Fe III. Typically not all lines are observed, but the typical line ratios are well-constrained \citep{francis91} and L$_\mathrm{BLR}$ can be calculated from the observed luminosity of a few lines \citep{celotti97}. Typical L$_\mathrm{BLR}$ for quasars are 10$^{43-46}$ erg/s and for BL~Lacs (mostly LSPs) 10$^{41.5-45}$ \citep[][]{celotti97}. In addition to the luminosity, one also needs to know the size, which is typically estimated by scaling from the disk luminosity \citep{ghisellini09}, which in turn is usually estimated from the UV data \citep[see e.g.][]{pian99}. For the dusty torus, the luminosities rarely come from direct observations because detection of a dust component in the IR SEDs of blazars has proven difficult owing to the dominance of the non-thermal component. The presence of a hot dust component has been inferred in the quasars 3C~273 \citep{wills89, soldi08}, PKS~1222+216 and CTA~102 \citep{malmrose11}. \cite{malmrose11} also calculated that the luminosity of the thermal emission in PKS~1222+216, $8\times10^{45}$ erg s$^{-1}$, is sufficient to supply the bulk of the seed photons for IC scattering if the emission region is within the radius of the dusty torus. The size of the dusty torus can be constrained from the theoretical considerations of the dust sublimation radius \citep[e.g.][]{nenkova08} and from reverberation measurements \citep{minezaki04,suganuma06}. These show that the inner radius of the dust emitting regions of AGN is two to three times smaller than the theoretical value, perhaps because only the largest dust grains persist in the inner torus. The size of the dusty torus is typically estimated to be $1-5$ pc and it scales with disk luminosity.

When one moves away from one-zone models, the number of free parameters for reproducing the observed SED increases. For these models it is even more important to be able to limit at least some parameters directly from the observations. 

\subsection{Hadronic models}\label{hadronic}

In the previous subsections we have neglected the hadronic models for producing the second peak of the SED, and only discussed inverse Compton scattering. But the same processes that accelerates the synchrotron emitting electrons to high energies are also expected to accelerate protons and nuclei. Proton blazar models \citep{mannheim93} were initially motivated by the search of sources of ultra-high-energy cosmic rays, not with a particular need to explain some observed features in the blazar SEDs or variability. 

In hadronic models the $\gamma$-ray emission can be proton synchrotron emission or photo-pion production and higher order processes with multi-pion production, followed by pion and muon decays that, in addition to $\gamma$-rays (from neutral pion decay), produce relativistic electrons, positrons and neutrinos. Also muon synchrotron emission might be an important process \citep{mucke03}. Efficient neutrino production in AGN jets requires high values of proton power, orders of magnitude higher than we see in leptons \citep[see e.g.][in the specific case of TXS0506+056]{cerruti19} or an extremely dense target photon field, which implies strong $\gamma$-ray absorption above the threshold for pair production \citep{waxman99}. 

The main challenge for hadronic models is that they require the power in relativistic protons to be in the range $L_{p}\sim 10^{47}-10^{49}$ erg s$^{-1}$ \citep[see e.g.][]{bottcher13}, in most cases dominating the total power in the jet. There are some differences between models though. For example, \cite{zech17} find that for two close-by HSPs, Mrk421 and PKS~2155-304, the energy carried by relativistic protons and the energy carried by the magnetic field are close to the equilibrium. The ratio between these two is a critical point as it also affects which of the channels, proton synchrotron, muon synchrotron, or cascading started by photo-pion production, would dominate the observed emission. 

\begin{figure}
\includegraphics[width=\hsize]{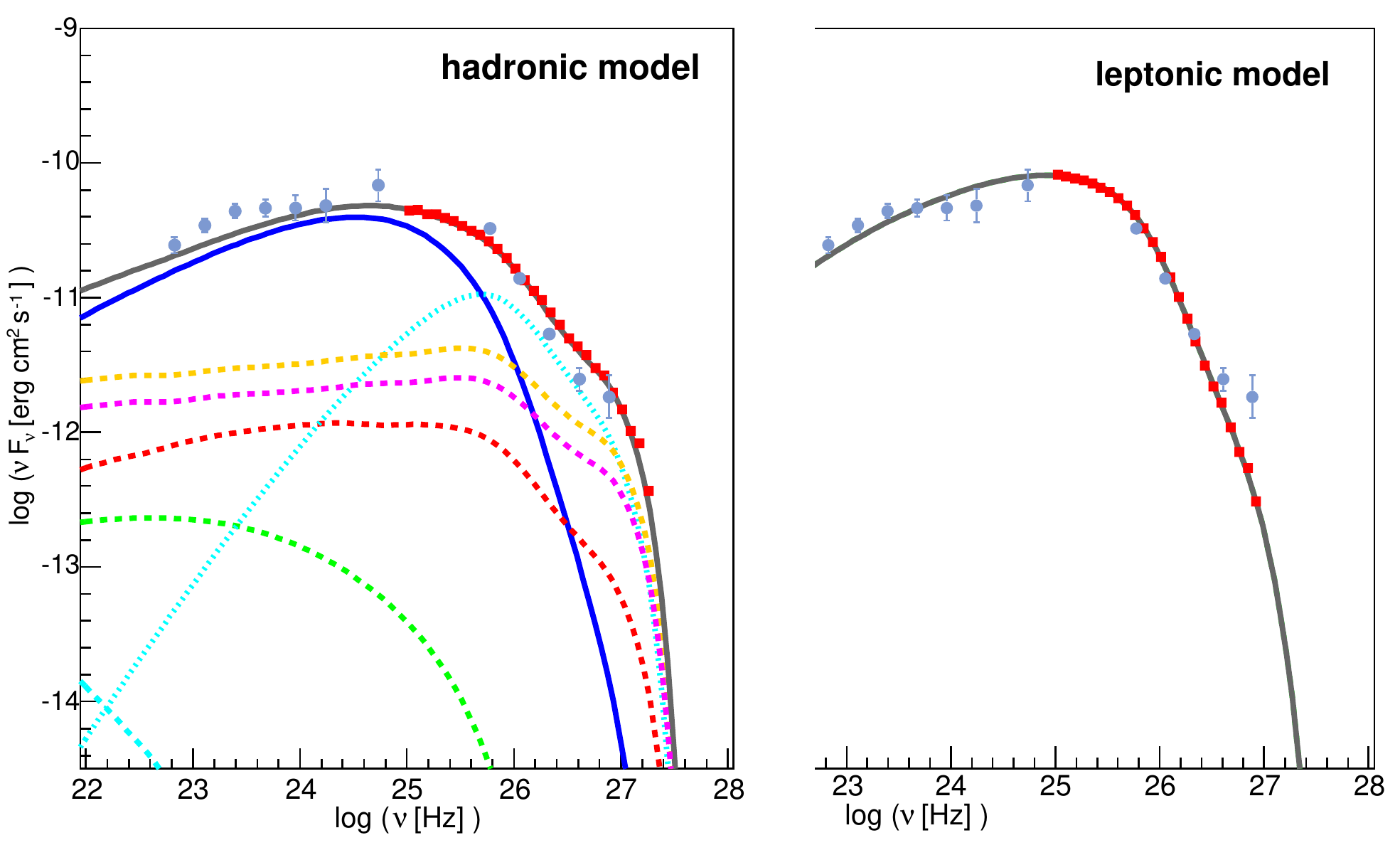}
\caption{Hadronic (left) and leptonic (right) $\gamma$-ray SED for PKS~2155$-$304. The red data points are simulated CTA data, demonstrating that "cascade bump" (see text) could be detectable with CTA. Figure taken from \cite{zech17}. Reproduced with permission from Astronomy \& Astrophysics, \copyright ESO.}
\label{fig:zech17}
\end{figure}

Proton synchrotron emission typically requires magnetic field strengths $B$ on the order of $10$\,G, and as this seems to be in conflict with the values derived from VLBI observations (see the previous section), also models requiring lower $B$ values have been investigated. As the magnetic field strength decreases, the peak frequency of the proton synchrotron emission decreases, and muon synchrotron emission and cascading becomes more important in the highest energies. The spectral hardening due to internal synchrotron-pair cascades, the "cascade bump", is possibly a distinguishable signature of hadronic $\gamma$-ray emission (see Figure~\ref{fig:zech17}). For it to appear, the ratio between the kinetic energy density of the relativistic protons and the energy density of the magnetic field has to be sufficiently large so that proton-photon interactions are non-negligible against proton-synchrotron emission \citep[see e.g.,][]{zech17}. 

In addition to the open question of the dominating channel in hadronic mechanisms, there is also ongoing discussion  for which sources the conditions for hadronic mechanisms and neutrino production are most favorable. First works considered FSRQs \citep{mannheim93}, but it was shown by \citet{sikora09} that the observed hard spectral indices observed in X-rays would be very hard, if not impossible, to reproduce if hadronic processes were in an important role in producing the second peak of the SED in FSRQs. This has been shown also in recent simulations \citep[e.g.][]{petropoulou15}. However, FSRQs seem to be the most promising sources of ultra-high-energy neutrinos due to the presence of external photon fields \citep{murase14}. 

The role of the hadronic processes in BL~Lac objects was evaluated in \cite{mucke01} and \cite{mucke03}. They found that in low-frequency peaked BL~Lacs photo-pion production and subsequent cascading, including synchrotron radiation by muons, are important processes. This also means that neutrino production is more efficient in LSPs.
They also found that, as in FSRQs, also in these sources the magnetic field should be on the order of 10\,G for the hadronic processes to dominate the second peak of the SED.
\cite{zech17} applied hadronic modeling to two of the most studied BL~Lac objects, Mrk~421 and PKS 2155$-$304 and found that for Mrk~421 muon-synchrotron
emission dominates, while proton-synchrotron radiation dominates in PKS~2155$-$304.
\citet{cerruti15} studied the extreme BL~Lac objects, with $\nu_C >1$ TeV and found that for magnetic field strengths of $<1$\,G the second peak of the SED is a sum of (leptonic) synchrotron self-Compton and synchrotron emission from muons and other cascading processes, while for higher magnetic field values the proton synchrotron would be in an important role. \citet{righi19} suggested that radiatively inefficient accretion flows could provide an important external photon field for neutrino production and that these fields would be much more luminous in LSP sources than in HSP sources, explaining why the two closest and brightest HSP sources, Mrk~421 and Mrk~501, have not been detected by IceCube. So, in summary, there seems to be a consensus that hadronic mechanisms are unlikely to dominate the second peak of the SED, but it might have a significant contribution in some BL~Lacs, but it is still under a debate, if it is LSPs, ISPs, HSPs or extreme HSPs that have highest hadroness (i.e. hadronic processes contribute most to the SED).

Finally, the case of TXS 0506+056 must be mentioned as several modeling efforts of its SED have been published lately and its classification has been discussed. It does not show strong emission lines nor extremely high or low synchrotron peak frequency \citep[it is ISP according to][]{ackermann11}, but it is more luminous than average ISP sources \citep{righi19}. \cite{ansoldi18},\cite{keivani18}, and \cite{cerruti19} all found that a physically consistent picture can only be found with $\gamma$-rays produced by inverse Compton processes, and high-energy neutrinos via a radiatively subdominant hadronic component.{\footnote{However, \citet{reimer19} pointed out that the neutrinos and $\gamma$-rays from TXS 0506+056 in 2014-2015 could not originate from the same emission region due to opacity effects, which is in apparent conflict with the above-mentioned modeling works.}} This is in agreement with all the discussed results above, and with the constraints on the magnetic field strength from VLBI observations. 

\subsection{Magnetic fields}\label{Bfields}
Magnetic fields are thought to play a large role in both launching and collimating the jets. Moreover, they have a significant role also in particle acceleration and thus the flaring of blazars. Therefore, understanding the magnetic field structure in the jets of blazars both on large scales (relevant to jet launching and collimation) and on small scales (relevant to particle acceleration) is of utmost interest. Magnetic fields can be studied by polarization observations because the angle of polarization, the electric vector position angle (EVPA), is related to the direction of the magnetic field, although relativistic effects may make the interpretation quite complex \citep{lyutikov05}. 

Multifrequency radio polarization observations of radio galaxies and quasars were already done in the 1960s, following the discovery of 8\% polarized radio emission in Cygnus A \citep{mayer62}. Later in the same year, \cite{cooper62} detected changes in the linear polarization as a function of wavelength squared, consistent with Faraday rotation (see also section \ref{faraday} below). These observations opened a new window for studying magnetic fields in radio galaxies and quasars. One of the first catalogues of linear polarization was published by \cite{morris64} where both the fractional polarization and EVPA behavior across wavelength was discussed with the conclusion that at least in some sources, the EVPA was perpendicular to the double-source direction. In some sources, signatures of depolarization, consistent with Faraday rotation were also seen, while others showed more complex behavior, possible indicating multiple polarized components. 

\subsubsection{Polarization variability}
Following the detection of total intensity variability in extragalactic radio sources \citep{dent65}, these sources were also seen to be variable in linear polarization \citep{aller67}. In the 1980s it was noted that during radio flares, the fractional polarization was seen to increase, which is a signature of magnetic fields getting more ordered, for example, due to shocks compressing an initially turbulent magnetic field \citep{laing80, hughes85}. Much of the work on polarization variability of blazars has been done at the University of Michigan Radio Astronomy Observatory, where more than a hundred blazars were monitored at 4.8, 8, and 14.5\,GHz until mid 2012. Studies of complete samples of quasars and BL~Lac objects have revealed that the differences in 4.8 and 14.5\,GHz polarization are mainly due to opacity effects \citep{aller99,aller03}.

The cm-band polarization observations in many sources are also consistent with transverse or oblique shocks \citep{hughes85,hughes89,hughes89b,aller14}. As discussed in \citep{hughes15}, these cm-band linear polarization observations can be used to independently constrain some of the fundamental jet parameters, such as the Lorentz factor and viewing angle of the source. All these cm-wavelength studies are consistent with shocks as the primary particle acceleration mechanism in the jets. However, because of the opacity in the jets at cm-wavelengths, in order to study the magnetic fields close to the black hole, and most relevant to jet launching and perhaps high-energy emission, one needs to go to shorter wavelengths where the jets are optically thin.

Optical polarization observations of blazars have also been conducted since the 1960s \citep[see][for a review]{angel80}. Unlike in normal galaxies, where the optical polarization is typically due to the scattering from dust, in blazars the optical polarization is also from synchrotron emission in the jets. In fact, one of the definitions of blazars includes that they must show optical polarization at a level higher than 3\% \citep{angel80}. In a study of more than 100 blazars in the RoboPol sample, \cite{angelakis16} find a clear trend in the optical polarization as a function of SED peak, with the HSP sources showing much lower polarization than the other blazar classes. Similar trends are also seen in the MOJAVE observations of the radio core polarization \citep{lister11,hodge18}, suggesting that there are differences in the magnetic field properties of the different types of objects (see below for further evidence when also the EVPA behavior is considered). 

As discussed in the comprehensive review by \cite{angel80}, the optical polarization was soon found to be variable with nightly variations \citep{kinman68}. The first smooth optical EVPA rotation was reported by \cite{kikuchi88} who observed a $120^\circ$ rotation in the blazar OJ~287. At the same time, the 10\,GHz radio EVPA was seen to rotate by $80^\circ$. They interpreted the rotation to be due to a shock traveling in a helical field, as suggested by the model of \cite{konigl85} that was inspired by the rapid EVPA swings observed earlier in the radio wavelengths \citep[e.g.,][]{ledden79, altschuler80}. Recently, these data were combined with the 4.8, 8, and 14.5\,GHz data from the UMRAO program, and several $180^\circ$ rotations were detected over a time period 40 years \citep{cohen18}. A model with two polarized components with counterrotating EVPA superposed on a steady polarized jet can explain many of these features. These kind of structures could be generated, for example, when forward and reverse pairs of fast and slow magnetohydrodynamic waves travel in a helical field \citep{nakamura10,nakamura14}.

\begin{figure}
\includegraphics[scale=0.33]{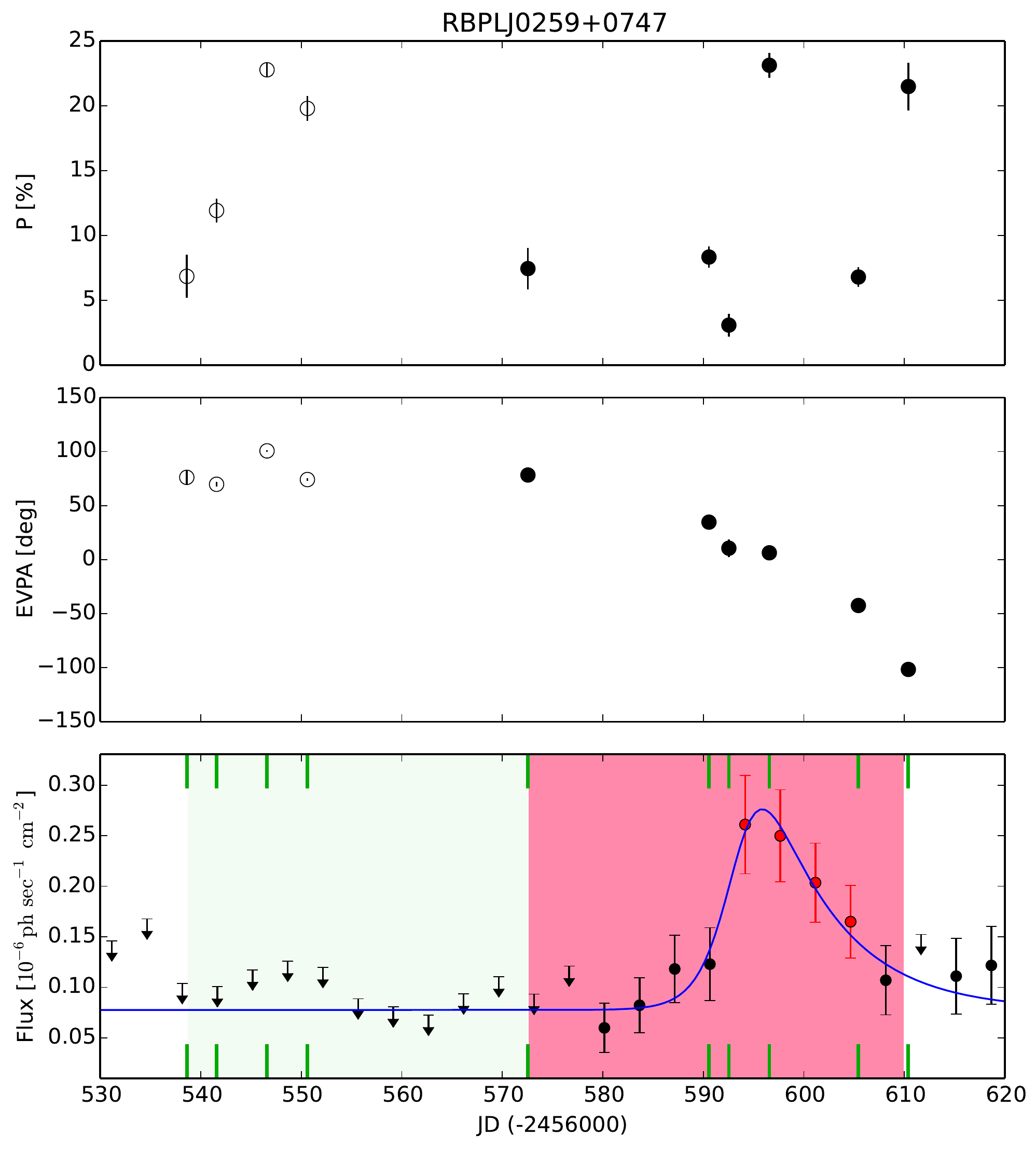}\includegraphics[scale=0.33]{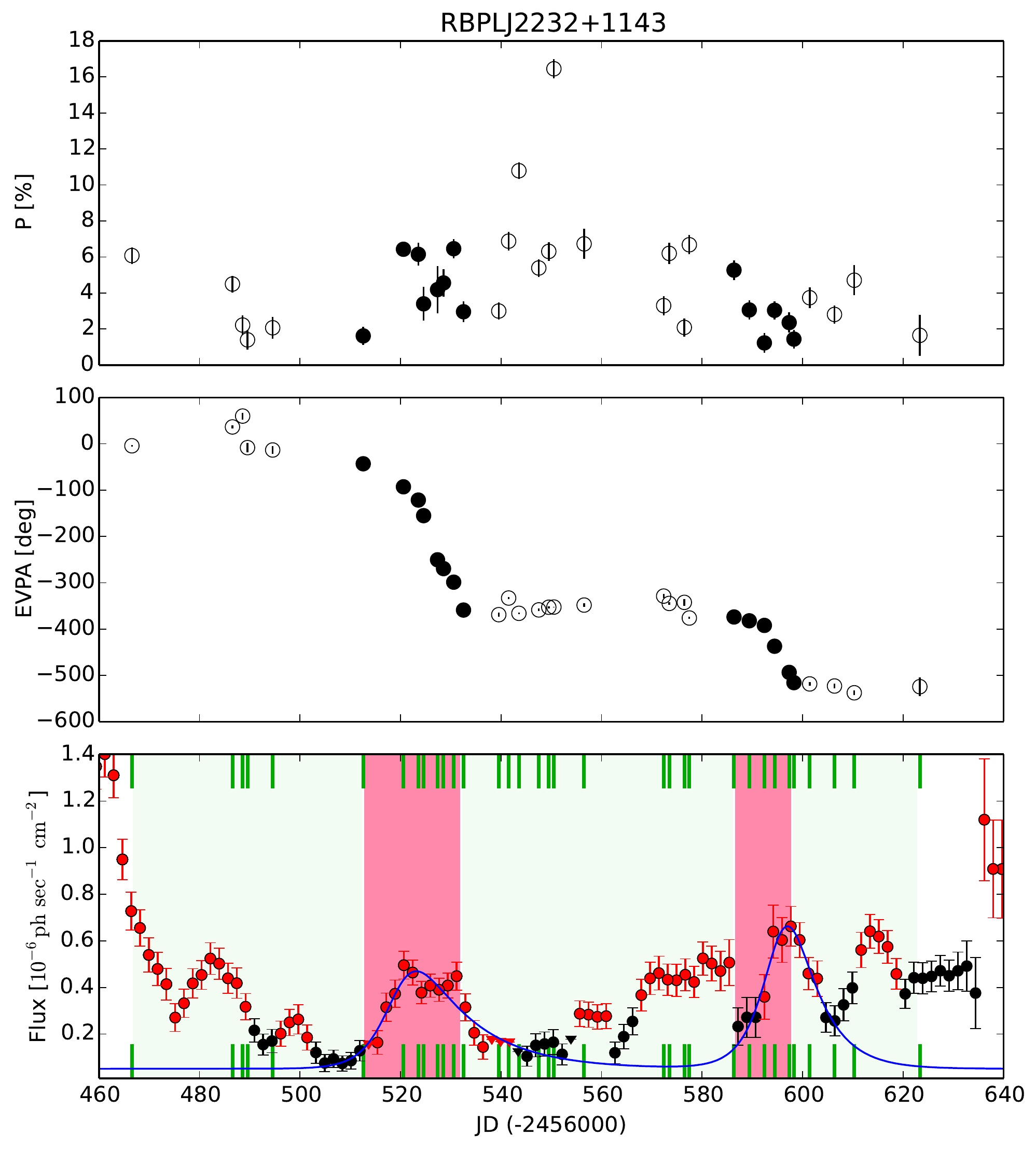}
\caption{Examples of two blazars observed within the RoboPol program. The top panels show the polarization degree, the middle panel the EVPA, and the bottom panel the $\gamma$-ray light curve observed by the {\it Fermi}-LAT. In the polarization plots, the filled symbols indicate periods when a rotation in the EVPA is seen. The shaded regions in the bottom panel show the same time period in the $\gamma$-ray light curve, indicating that in both sources the rotations happen during $\gamma$-ray flares. Both objects also show very fast variability in the polarization degree. Figure adapted from \cite{blinov15}.}
\label{fig:blinov}
\end{figure}

Alternative interpretations for the EVPA rotations included a simple two-component model \citep{bjornsson82} where the rotation was due to a change in the fluxes of two components with different polarization properties. \cite{bjornsson82} also discuss a relativistic model for the polarization variations, where the rotations are due to an aberration effect when a relativistically moving polarized source travels down the jet. This would require symmetry in the underlying magnetic field. On the other hand, based on radio observations, \cite{jones85} suggested that the rotations could be explained by random walks in a largely turbulent magnetic field. 

The EVPA rotations became a hot topic again when \cite{marscher08} observed an optical EVPA rotation of $240^\circ$ in the source BL~Lacertae. The rotation was coincident with flaring in X-ray, optical and TeV energies, and a passage of a new superluminal component through the 7\,mm VLBI core. \cite{marscher08} interpreted this rotation to be due to a polarized component traveling in a helical magnetic field. Since the launch of the {\it Fermi} satellite, the connection between optical rotations and gamma-ray flaring became more evident \citep[e.g.,][]{marscher10,abdo10}. 

A major improvement in the statistics of optical EVPA rotations and polarization variability was achieved through the RoboPol program, who monitored a large number of gamma-ray detected and non-detected objects in R-band optical polarization \citep{pavlidou14}. In \cite{blinov18} the latest results, summarizing the observations from the three years of RoboPol operations in 2014--2016, are discussed. RoboPol detected 40 rotations in 24 objects, which were all found to be coincident with gamma-ray flares detected by {\it Fermi} (see Fig.~\ref{fig:blinov} for examples). This suggests that the magnetic fields and flaring are tightly connected. This is further supported by the observation that $\gamma$-ray detected objects have higher optical polarization than the non-detected sources \citep{pavlidou14,angelakis16}.

In addition to showing EVPA rotations, some objects are seen to exhibit very stable EVPA over many years, indicative of an ordered and stable magnetic field component \citep[e.g.,][]{angel78,hagen80,jannuzi94,villforth10,hovatta16}. In some objects, the preferred angle is seen only intermittently \citep[e.g.,][]{hagen80,hagen02,villforth10}, suggesting that this could be related to an underlying stable component being confused with a varying polarized components, as suggested, for example,  in the model by \cite{bjornsson82}. The stability of the EVPA seems to also be connected to the SED peak of the sources. When the stable EVPA is compared with the position angle of the parsec-scale jet obtained through VLBI observations, in quasars there is no clear connection \citep{lister00,jorstad07,angelakis16}, while in BL~Lacs and especially in high synchrotron peaking BL~Lac objects the EVPAs are more stable \citep[e.g.,][]{angelakis16} and they are more aligned with the jet position angle \citep{jorstad07,hovatta16}. 

This could be explained with a model where the emission is due to a shock compressing a helical or toroidal magnetic field \citep{angelakis16}. In HSP sources, the optical emission originates from lower energy electrons than in LSP sources, and in this model, the emission in HSPs would come from a larger volume, resulting in an overall lower polarization fraction, as observed in the RoboPol sample \citep{angelakis16}. In addition, the emission would be dominated by the stable helical or toroidal field component, resulting in a stable EVPA aligned with the jet direction. An alternative explanation is a spine-sheath structure \citep[e.g.,][]{ghisellini05}, where the optical and radio emission would come from the outer, slower sheath layer that is dominated by a helical or toroidal field, again resulting in a more stable EVPA aligned with the jet direction. This would also be consistent with the slower apparent speeds observed in HSP sources \citep[e.g.][see also Fig.~\ref{fig:lister19}]{piner18,lister19}.

Despite the recent progress in observations by RoboPol and others \citep[e.g.,][]{jermak16}, it is still unclear if there is a single dominant magnetic field configuration responsible for all flares and rotations in all sources, or if multiple mechanisms are at play. In the last few years, there have been several new theoretical studies on the underlying mechanism of the rotations, ranging from turbulence to shocks to magnetic reconnection \citep[e.g.,][]{marscher14, hughes15, zhang15, zhang16, zhang18}. For details of these models and their comparison, see the recent review by \cite{bottcher19}. Some detailed observational studies, such as \cite{kiehlmann16} on the blazar 3C~279, emphasize the need for very good sampling as spurious rotations are easily seen due to the $n\times\pi$ ambiguity in the EVPA, when the sampling is inadequate. Thus, there are still major open questions in regards to the magnetic field structure and the particle acceleration mechanism in the flaring regions. 

One way to try to answer these questions is through combined radio, millimeter and optical polarization observations \citep[e.g.,][]{rudnick78,gabuzda94,lister00,jorstad07}. Simply detecting similar polarization degree and EVPA is not enough to establish a co-spatial origin of the emission, as this could also be due to the magnetic field between the regions being uniform \citep{gabuzda94}. Thus, one needs to observe simultaneous variability in the different wavelengths to establish a common origin for the emission. \cite{jorstad07} studied 15 objects at optical, 1\,mm, 3\,mm and 7\,mm wavelengths, where the 7\,mm observations (43\,GHz) were obtained with the VLBA and included spatial information about the polarization structure. Their study was one of the first multi-epoch studies where the same objects had been monitored for 3 years. Although they found a good correspondence between the 43\,GHz core and optical polarization, perhaps surprisingly, the connection between the higher mm-band wavelengths and optical was less clear. They also noted that while in some sources the EVPAs between different wavelengths agreed well, in others there was no clear connection. 

Although detailed studies of individual objects including both VLBI and optical observations have allowed to construct models that fit the data well \citep[e.g.][]{marscher08, darcangelo09}, we still lack a clear understanding on how this relates to the type of the blazar, and if there can be different mechanisms at play even in the same object. It is likely that both shocks and turbulent processes play a role, complicating the picture \citep{jorstad13}. Only by obtaining densely sampled {\it long-term} data of a large sample of objects can this be achieved.

\subsubsection{Faraday rotation}\label{faraday}
Faraday rotation is a propagation effect, where the intrinsic polarization of a synchrotron source is altered due to magnetized plasma between the source and the observer. Most notably, the observed EVPA is rotated with respect to the intrinsic one. This rotation is wavelength dependent ($\lambda^2$) and proportional to the line-of-sight component of the magnetic field and the electron density in the intervening plasma. As discussed earlier, the first detection of Faraday rotation in a radio galaxy was for Cygnus A \citep{cooper62}. By studying the spatial distribution of Faraday rotation across Cygnus A, \cite{gardner63} suggested that the observed Faraday rotation is either due to magnetic fields in our own Galaxy, or near Cygnus A. 

Since the 1960s, there have been numerous studies on the Faraday rotation in radio galaxies and blazars. Especially after the VLBA started operations in full-polarization mode in 1994, it has been possible to study the Faraday rotation distributions in a large number of blazars \citep[e.g.,][]{zavala03,zavala04,hovatta12}. These studies have shown that the Faraday rotation is typically higher in the core (on the order of $\sim10^3$\,rad/$m^2$) than the jet ($\sim10^2$rad/$m^2$), indicative of higher electron density and/or magnetic fields closer to the nucleus of the blazar. FSRQs are also seen to show higher Faraday rotation than BL~Lac objects, possibly indicating differences in the plasma surrounding these different blazar types.  

\begin{figure}
\includegraphics[width=\hsize]{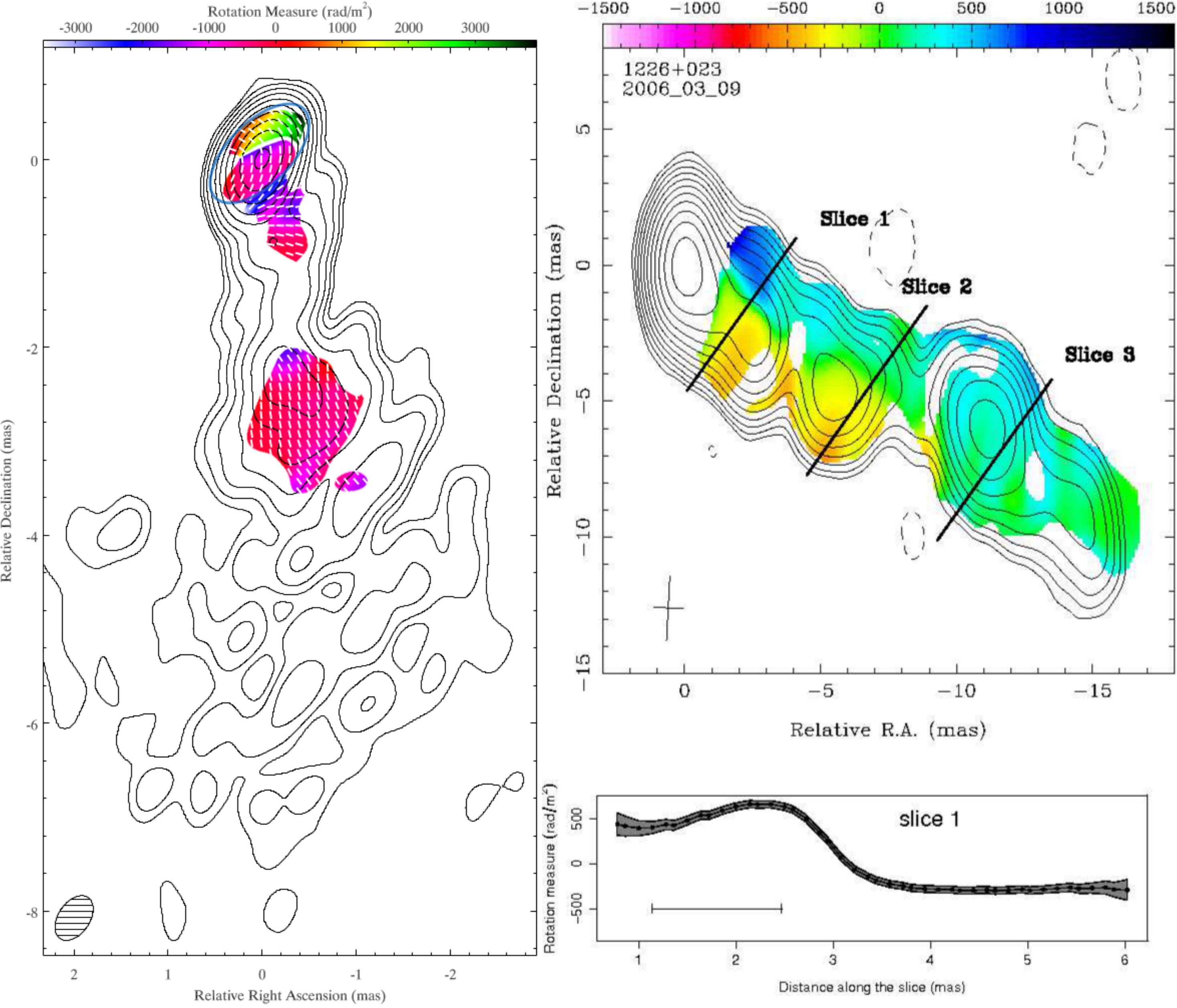}
\caption{Left: Rotation measure map of BL~Lac between 15, 22, and 43\,GHz, where the 22\,GHz image includes data from RadioAstron. The contours show the total intensity of the source, while the color scale indicates the amount of Faraday rotation. White ticks over the image show the Faraday-corrected EVPA direction. A gradient in the rotation measure is clearly seen over the core region, which is indicated by a circle. Figure taken from \cite{gomez16}. Right: Rotation measure map of 3C~273 at $8-15$\,GHz frequencies. Contours indicate the total intensity, and the color scale show the amount of Faraday rotation. A clear gradient is seen transverse to the jet direction over the entire length of the jet. The lower panel shows the rotation measure values over Slice 1 indicated on the image. Figure adapted from \cite{hovatta12}. Figures reproduced by permission of the AAS.}
\label{fig:rm}
\end{figure}

One useful property of Faraday rotation is that because the electron density is always positive, the sign of the rotation measure indicates the direction of the line-of-sight magnetic field, with a positive rotation measure for a magnetic field coming towards the observer. This lead to the suggestion that one could detect possible toroidal magnetic field structures by observing a gradient in the rotation measure transverse to the jet direction \citep{laing81,blandford93}. The first such gradient was detected in the FSRQ 3C~273 by \cite{asada02} in VLBA observations between 5 and 8\,GHz, who interpreted the gradient to be due to a sheath surrounding the relativistic jet. The gradient was later confirmed by \cite{zavala05} and \cite{hovatta12} in VLBA observations at 8 to 15\,GHz frequencies (see Fig.~\ref{fig:rm} right panel). If the gradient is due to a helical field in or around the jet, the total intensity and polarization profiles transverse to the jet should also be asymmetric \citep{clausen-brown11}, which seems to be the case for 3C~273 \citep{hovatta12}.

Since the first detection by \cite{asada02}, there have been many more claims of Faraday rotation measure gradients in blazars \citep[e.g.,][]{gabuzda04,gabuzda17}, however, there are not many objects that are well enough resolved in the direction transverse to the jet so that the asymmetries in total intensity, polarization and Faraday rotation could be studied in detail. In the highest angular resolution observations of RadioAstron, there is a clear change in Faraday rotation around the core of BL~Lac (see Fig.~\ref{fig:rm} left panel), which is also interpreted as a signature of a helical field threading the inner jet \citep{gomez16}.

General relativistic magnetohydrodynamic simulations by \cite{broderick10} and \cite{porth11} have also been able to reproduce many of the observed properties of Faraday rotation in pc-scale jets, including gradients transverse to the jet. In the model of \cite{broderick10}, the Faraday rotation originates from a region connected with the jet, such as a sheath, but they also caution against making strong conclusions based on transversely unresolved jets, as they show how large deviations from the true values can originate when the simulated jets are convolved with typical beam sizes of VLBA observations. Thus, it is still unclear whether all jets possess such magnetic field structures, or if this is a property of a small number of special objects. Better spatial resolution in the transverse jet direction would be needed to achieve this, which will hopefully be possible with the advent of the next generation sensitive VLBI arrays, and more high resolution observations such as the ones provided by RadioAstron \citep{gomez16}.

Most of the studies of Faraday rotation, and especially of gradients, have been done with the VLBA at cm wavelengths. This means that the scales probed by the observations are parsecs to hundreds of parsecs away from the black hole, after accounting for jet opacity and projection effects. The highest frequency VLBA can reach is 86\,GHz, which allowed \cite{attridge05} to study the polarization structure of 3C~273 at 43 and 86\,GHz. They detected a change in the EVPA at the two frequencies, which corresponds to a gradient of at least 20~000\,rad$/m^2$, indicating that, at regions closer to the black hole, the electron density and magnetic field strength further increases. In a conical jet under equipartition, the Faraday rotation measure is expected to follow a relation $|$RM$|\propto \nu^a$, where the value of $a$ depends on the power-law change in the electron density $n_e$ as a function of distance $r$ from the black hole, $n_e \propto r^{-a}$ \citep{jorstad07}. The value of $a$ is seen to vary depending on the source, but the median values are typically around 2, consistent with the Faraday rotation occurring in a sheath around a conically expanding jet \citep[e.g.][]{jorstad07, osullivan09,kravchenko17}. This means that with higher mm-band frequencies, one could be able to detect even higher rotation measures. 

A new avenue for detecting extreme rotation measures is offered through observations with ALMA. The first detection of an extreme Faraday rotation of $>10^7$ rad$/m^2$ was for the lensed quasar PKS~1830-211 \citep{marti-vidal15} at 230 and 345\,GHz frequencies. At the redshift of the target, $z=2.5$, this corresponds to a Faraday rotation of $>10^8$~rad/$m^2$ indicating extreme magnetic fields or electron density in the region where the rotation orginates. Interestingly, in observations taken at even higher frequency of 650\,GHz (corresponding to 2.3\,THz in the source frame) two years later, the observed rotation measure is much lower, only on the order of a few times $10^5$~rad/$m^2$ \citep{marti-vidal18}. Recently, very high rotation measure of $5\times10^5$rad$/m^2$ was also observed in the quasar 3C~273 over the ALMA 1.3\,mm band, which is consistent with a sheath surrounding a conical jet, when compared to lower frequency observations \citep{hovatta19}. These high values are also consistent with the simulations of \cite{porth11}, who find that in the millimeter range, the core Faraday rotation values can reach up to $10^6$\,rad/$m^2$.

The observations of varying rotation measure in PKS~1830-211 by \cite{marti-vidal18} show that the conditions in the jet launching region of AGN change over time, similar to what is seen in the mm-band Faraday rotation observations of Sgr A*, where the variations are thought to occur due to changes in the turbulent accretion flow \citep{bower18}. Only by obtaining more such observations of multiple AGN, can we establish whether this is also what is the cause of the variations in the supermassive active black holes. 

Recent simulations by \cite{moscibrodzka17} showed that, in case of M87, the polarized emission and Faraday rotation most likely originates in the forward jet instead of in the accretion flow, as typically assumed for low-luminosity galaxies \citep[e.g.,][]{plambeck14,kuo14}, complicating the picture further. Simulations of objects with efficient accretion are also required to extend the comparisons to high-luminosity objects, such as blazars. 

\subsubsection{Circular polarization}\label{cp}
Another observational way to probe the magnetic fields and also particle composition in blazars is through circular polarization observations. Circular polarization can either be intrinsic due to synchrotron radiation, or produced through Faraday conversion of linear to circular polarization \citep{komesaroff84}. While intrinsic circular polarization probes the magnetic field structure of the jets, the Faraday conversion is dominated by the low-energy particles in the jet, giving means to study, for example, the low-energy cutoff of the electron spectrum (see also section \ref{sedparameters}), and the particle composition in the jets \citep[e.g.,][]{homan99,beckert02}.

Circular polarization is much weaker than linear polarization, typically only some fractions of a percent \citep[e.g.,][]{homan06} up to a few percent \citep{homan04}, which makes it challenging to detect. 
While some sources show stable signs of the circular polarization over decades \citep[e.g.,][]{homan01,homan18}, especially at lower \citep{aller03b} and higher \citep{thum18} frequencies, the variations can be more erratic due to changes in jet opacity or the higher frequency observations probing smaller, possibly more turbulent, length scales of the magnetic fields \citep{homan18}. There also seems to be a frequency-dependence of the magnitude of circular polarization, with higher frequency observations often showing higher circular polarization values \citep{vitrishchak08,thum18}, which disagrees with expectations for simple homogeneous component models \citep[see][for a review]{wardle03}. 

It is, thus, still unclear what is the primary production mechanism of circular polarization in the jets, and as usual, detailed multifrequency studies reveal a complex picture. For example, using ATCA observations between 1 and 10\,GHz, \cite{osullivan13} constructed a circular polarization spectrum for the source PKS~2126$-$158, which together with linear polarization measurements was consistent with Faraday conversion from linear to circular polarization. On the other hand, \cite{homan09} found that while in the inhomogeneous jet base of 3C~279 the circular polarization was most likely intrinsic to synchrotron emission, the spatially resolved homogeneous components in the jet were more consistent with Faraday conversion, indicating that both processes may play a role at the same time. This, of course, complicates any studies where the emission regions cannot be spatially resolved, meaning that only by observing multiple sources with high spatial resolution at multiple frequencies, can we truly understand the origin of circular polarization and its role in the magnetic field structure and particle composition of the jets.

Circular polarization also offers an independent way to estimate the magnetic flux carried by the jet. If the jet is magnetically launched from the black hole ergosphere, the net magnetic flux is a conserved quantity so that the magnetic flux observed in the jet equals the magnetic flux at the central engine \citep{blandford77}. By using circular polarization observations to measure the fraction of poloidal magnetic field in the core of 3C~279, \cite{homan09} was able to estimate the magnetic flux of the jet. This value of observed magnetic flux can be related to the jet launching and accretion models, giving insight into the accretion disk structure and the relevance of magnetic fields in jet launching, as was done in \cite{zamaninasab14}. They found the disk luminosity and magnetic flux of the jet to be tightly correlated in a sample of 76 objects, in support of the magnetically arrested disk models \citep{narayan03,tchekhovskoy11,mckinney12}. In the work of \cite{zamaninasab14}, the magnetic flux was estimated using the magnetic field strength from core-shift measurements of \cite{pushkarev12}, and the circular polarization observations thus provide an independent way of measuring the magnetic flux and confirming the results.

\subsection{Structured jets}\label{structure}

\begin{figure}
\includegraphics[scale=0.45]{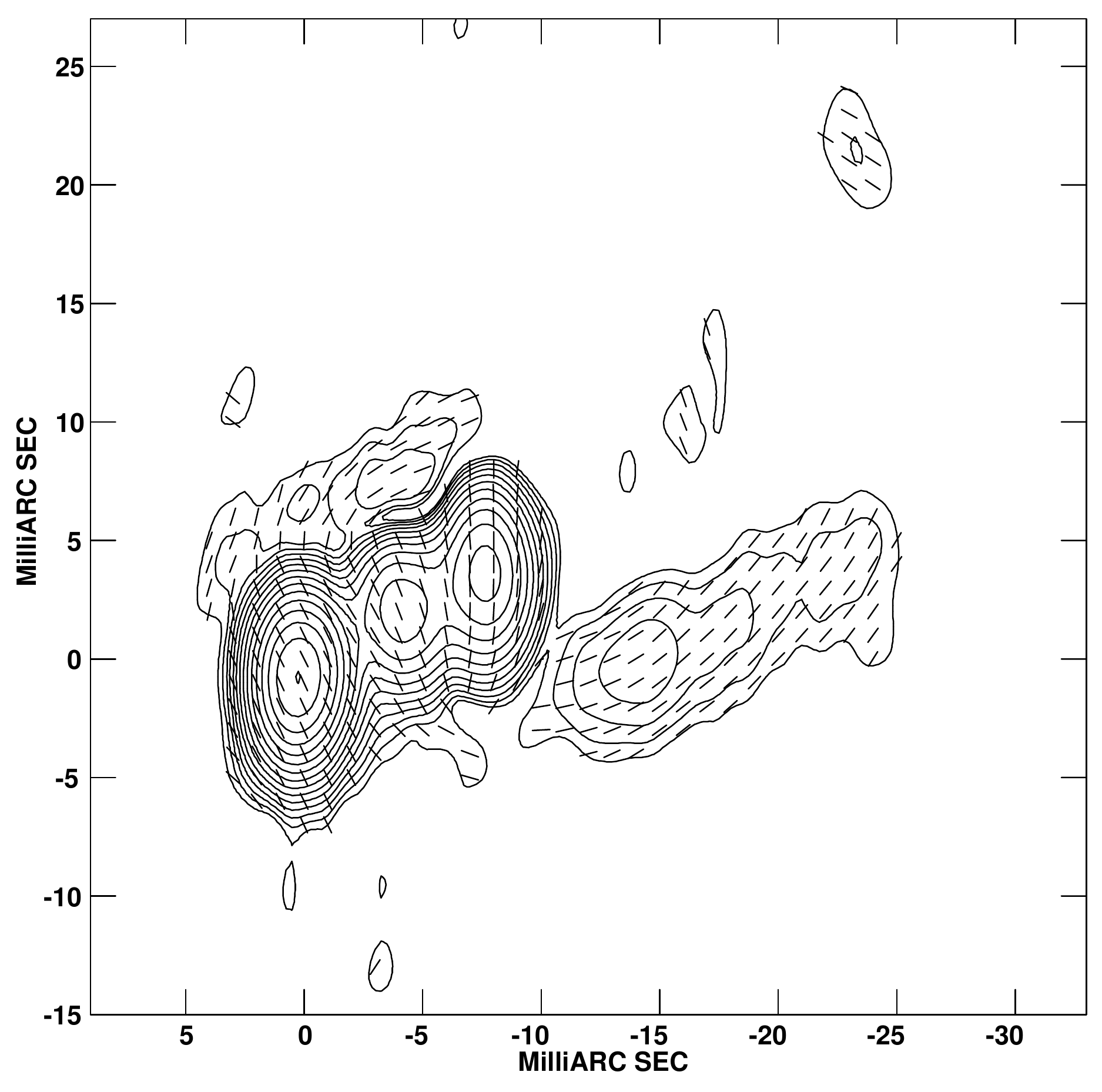}
\caption{5\,GHz VLBA image of the FSRQ 1055+018 by  \cite{attridge99}. The contours indicate the linear polarization intensity of the source while the ticks show the magnetic field direction (EVPA$+90$ degrees). The magnetic field in the inner jet is clearly different from the outer region, indicating a spine-sheath structure of the jet. Reproduced by permission of the AAS.}
\label{fig:attridge}
\end{figure}

Observational evidence for structured jets in pc-scale jets of blazars was first discovered in polarimetric VLBA observations of the FSRQ 1055+018 by \cite{attridge99}, who saw a clear difference in the polarization direction of the inner jet, where the polarization angle was predominantly perpendicular to the jet versus in the outer layer where it was parallel to the jet (see Fig.~\ref{fig:attridge}). Similar polarization structures have since been observed in several other sources as well \citep[e.g.,][]{pushkarev05,gabuzda14}. Another indication of spine sheath structures is the limb brightening of the jet, where the edges of the jet appear brighter than the central spine. These kind of structures have also been observed in several high-resolution observations of radio galaxies and blazars on parsec scales \citep[e.g.,][]{giroletti04,piner09,nagai14,giovannini18}.

Direct evidence for different velocity structures in the well-resolved jets of 3C~273 and M87 were obtained by \cite{mertens16} and \cite{mertens16b} who analyzed a set of VLBA observations using a wavelet-based method, suitable for tracing features in jets without distinct blobs, such as the pc-scale jet of M87. This novel method is very promising for detecting moving velocity structures especially in HSP sources that typically do not show many distinct blobs \citep[e.g.,][]{piner18}, but it requires the jet to be well resolved transversely, which limits its use until more sensitive VLBI arrays become available.

The first theoretical model describing a two-flow model where an outer, mildy-relativistic wind, is ejected from the accretion disk, and an inner relativistic electron-positron jet forms within it was by \cite{sol89}. In \citet{ghisellini05} the structured jet was described as a cylinder (spine) surrounded by a hollow cylinder (layer), with the spine and the layer moving at different Lorentz factors. The very important consequence of this model is that the seed photons relevant for the scattering process are produced not only by the spine (layer) electrons, but also by the layer (spine) ones. \citet{ghisellini05} also pointed out that the slow layer also produces a large amount of GeV radiation, which remains visible even at large viewing angles, making the radio galaxies bright in GeV and TeV ranges, which is indeed what we are seeing with the current generation of instruments. 

These types of two-zone models are now gaining more popularity, as it is difficult to explain the very high and low energy emissions with single-zone models (see also Section \ref{seedphotons}). However, spine-layer (or spine-sheath as they are also called) models seemed to not be a solution to all observed properties, in particular to the bright $\gamma$-ray flares of some FSRQs without a high amplitude counterpart in lower energies (also called "orphan" flares). The discrepancy between the amplitudes is unexpected if the same electrons that are responsible for the optical synchrotron emission upscatter photons resulting in inverse Compton emission seen in the $\gamma$-ray energies. \citet{nalewajko14} showed that a sheath of plasma surrounding the jet would not be able to provide sufficient numbers of seed photons to produce the orphan flares seen in PKS~1510$-$089, without the sheath itself outshining the observed flux. 

This particular discrepancy of the lower amplitudes of optical flares compared to $\gamma$-ray flares was also addressed by the simple model of \cite{sikora16}, where it was shown that in a jet, where the inner spine moves at a higher velocity than the outer sheath, if the jet is viewed at a small angle, the inverse Compton emission is dominated by the emission from the spine, while the synchrotron emission originates in the slower sheath. In fact, in their model, the synchrotron emission from the spine was never dominant. Depending on the viewing angle, a different contribution of the spine and sheath emission is responsible for the high-energy emission. However, as stated by the authors themselves, this very simplistic model may not be adequate if the velocity structure of the jet is smooth compared to a step-like change in the Lorentz factor of the jet. 

\citet{macdonald15} also presented numerical calculations of the time-variable emission as a plasmoid propagates relativistically along the spine of a blazar jet and passes through a synchrotron-emitting ring. The ring represents a shocked portion of the jet sheath. It creates a very localized source of seed photons that are inverse-Compton scattered by the electrons in the moving blob. They demonstrated that this "ring-of-fire" model can create an orphan $\gamma$-ray flare as the blob passes through the ring. \citet{macdonald15} also compared their calculations, which reproduced the SED and light curves of the $\gamma$-ray flare well, with the observed size of the sheath from VLBI observations and found some factor of 5 discrepancy (the observed sheath being larger in PKS~1510$-$089).

Recently, \cite{vuillaume18} modeled successfully the quiescent emission of the FSRQ 3C~273 using a two-flow model where the plasma is accelerated through the Compton rocket effect. While the model was able to reproduce the observed SED of the source, the final bulk Lorentz factor of the flow was only $\Gamma=2.7$, much smaller than inferred from VLBI observations of the source \citep[e.g.,][]{lister09}. These attempts show that there is still room for improvement on the modeling side, to make the models match all available observational constraints.

\subsection{Periodicities}\label{periodicity}
Based on galaxy evolution through mergers, it is expected that active galaxies harbor binary supermassive black holes that may result in observed features such as jet precession visible in high angular resolution observations, or periodicities in their light curves \citep{begelman80}. One of the most clear and well-studied examples is the $\sim 12$ year periodicity in the optical light curve of OJ~287, which was interpreted to be due to the orbital period of $\sim 9$ years (in the source frame) of two supermassive black holes \citep{sillanpaa88}. What is special in the case of OJ~287 is that the periodicity holds even after $>100$ years of observations, and new outbursts follow the expectations of general relativistic models \citep[e.g.][]{valtonen16}.

Throughout the years, there have been several claims for periodicities in many other blazars both in radio and optical wavelengths, but in most cases the periodicities are not persistent when long time series are analyzed \citep[e.g.,][]{hovatta08}, or they can be better explained with a stochastic red noise process \citep[e.g.][]{vaughan16}. These kind of quasi-periodic oscillations (QPOs) on time scales of a few hundred days to some years could be due to e.g., tilted accretion disks \citep[e.g.,][]{liska18} also seen in X-ray binaries \citep[e.g.,][]{stella98}. With the advent of wide-field optical transient factories, the number of candidate binary systems has increased \citep[e.g.][]{graham15,charisi16}, but the light curves are typically still too short to reliably establish whether the periodicity is persistent or if it falls into a category of a QPO. 

Recently, periodic flares on time scales of $\sim2.2$ years were also seen in the $\gamma$-ray light curve of PG~1553+113 \citep{ackermann15b}. Periodicity on a similar time scale was also seen in the optical light curve, while in radio the flares appeared more erratic, and the possible periods did not coincide (which is not necessarily unexpected if the radio emission originates in a different part of the jet). The $\gamma$-ray light curve by {\it Fermi} was only 6.9 years long meaning that the periodicity could not be confirmed with high significance, and it could still fall into the category of a QPO, a longer time series is needed to confirm its nature. While it is interesting that the period was seen in both $\gamma$-ray and optical light curves, one should remember that the same particles are responsible for the emission in both wavelength regimes, meaning that also a stochastic process could be responsible for the variations. 

Ultimately, in addition to careful analysis to exclude the possible stochastic nature of the variations, one needs additional observational evidence before claiming that any (quasi)-periodicity is due to a binary black hole system. Ideally, this should come from a detection of a double nucleus or jet also in a blazar, as is already seen in some radio galaxies \citep{rodriguez06,kharb17}.  The challenge is of course the required angular resolution. In OJ~287 the expected binary separation is on the order of 0.1\,pc \citep{sillanpaa88}, which corresponds to $\sim20\mu$as taking into account the redshift of 0.306 \citep{stickel89}. This separation may be detectable with the Event Horizon Telescope \citep{doeleman09}, but for any binaries with shorter periods the separation is typically too small to be spatially resolved without space-based mm-band interferometry. 
Another way to potentially discover supermassive binary black holes, especially during their mergers, will come through the Laser Interferometer Space Antenna (LISA, \citealt{lisa17}) gravitational wave observatory that is sensitive to mergers of massive black holes (see also section \ref{lisa} below). 

\section{Outlook} \label{outlook}
There are several upcoming new instruments both on the ground and in space that will allow us to answer some of the remaining open questions in blazar science, and hopefully also result in many unexpected discoveries. Below we have gathered a limited subset of the major instruments that are expected to be operational in the next two decades.

\subsection{Event Horizon Telescope}
The Event Horizon Telescope (EHT,\citealt{eht2}) is an interferometer operating at 1.3\,mm wavelength. At the moment, it consists of telescopes in the United States, Chile, Spain, Mexico, South Pole, and Greenland,  with the phased ALMA array in Chile included as a single element in the array. As discussed in Section~\ref{highres}, the main goal of the project has been to image the shadows of the black holes in the center of our Milky Way, and in the nearby radio galaxy M87. The first image of the black hole's shadow in M87 was very recently published by the Event Horizon Collaboration \citep{eht1}.

With the demonstrated angular resolution of $20\mu$as in the images \citep{eht3}, it will be possible to study blazars in superb detail. Especially with the inclusion of polarimetry, it is possible to obtain essential information on the jet launching regions and their magnetic fields. Some blazars have already been observed with the full EHT array in 2017 and 2018 with ALMA included in the array, and the first results of the blazar 3C~279 were shown in \cite{kim20}. The array also continues to be offered as part of ALMA observing cycles so that we can expect several discoveries and detailed studies in the next years.

\subsection{X-ray polarimetry}
One of the next major breakthroughs in blazar science is expected to come through X-ray polarimetry. Especially in high synchrotron peaking sources, the emission in X-ray energies is synchrotron emission from high-energy electrons, meaning that there is potential that X-rays probe regions closer to the particle acceleration site than optical bands \citep[e.g.,][]{tavecchio18}. This allows one to potentially study the particle acceleration mechanism in blazar jets. As discussed in \cite{tavecchio18}, magnetic reconnection, which requires turbulent magnetic field structures, may not produce any polarized emission, while shock compression would result in a higher polarization fraction in X-ray bands than in the optical, especially in the HSP sources. As the authors themselves state, these conclusions should be re-visited with models accounting for the large-scale structure in the jets. 

The main benefit from X-ray polarimetry is determining whether the high-energy emission in LSP and ISP sources is due to synchrotron self-Compton or external Compton processes. Based on the properties of inverse Compton scattering, if the emission is due to EC the polarization fraction is much lower than the synchrotron polarization if the external photon field is isotropic (as usually assumed for blazars) \citep[e.g.,][]{bonometto70,krawczynski12}, while in case of SSC the X-ray polarization is expected to be about half of the synchrotron polarization \citep[e.g.,][]{liodakis19}. Moreover, X-ray polarization could be used to distinguish between leptonic and hadronic models \citep[e.g.,][]{zhang13}.

The most promising upcoming instrument for detecting X-ray polarization in blazars is the Imaging X-ray Polarimetry Explorer (IXPE) mission \citep{weisskopf16}, expected to launch in 2021. Based on calculations in e.g., \cite{tavecchio18}, IXPE should be able to detect 30\% polarization in bright X-ray blazars, such as Mrk~421 and Mrk~501 in less than 1\,ksec. This calculation is supported by the detailed study of \cite{chakraborty15} who estimated the prospects for blazar observations with several existing and upcoming X-ray missions, concluding that the balloon-based experiments will not yet have sufficient sensitivity for blazar detection in a reasonable time. Recently \cite{liodakis19} used a multi-zone jet model and earlier X-ray and optical observations to estimate what type of sources will be detected by IXPE. They conclude that the most likely objects to be detected are HSP sources, but also ISP sources will be accessible by IXPE. On the other hand, if LSP sources are detected, it either requires very long exposure times, or the origin of the radiation to be e.g., proton synchrotron emission.

\subsection{The Cherenkov Telescope Array}\label{cta}

The Cherenkov Telescope Array (CTA){\footnote{\url{http://www.cta-observatory.org}}} will be the first open VHE $\gamma$-ray observatory. It will consist of $\sim$100 Cherenkov Telescopes of three different sizes at two sites (North: La Palma, Spain and South: Paranal, Chile). It will provide an order of magnitude improvement in sensitivity and it will also extend the observable energy range significantly compared to the current generation VHE experiments. For blazar observations, pushing the energy threshold towards $\sim$20\,GeV is particularly important as most of the blazars show soft spectra in the VHE $\gamma$-ray range. The construction of the telescopes has already started at La Palma, and CTA should be operational by 2024. 

Blazars have a significant role in the key science programs of the CTA consortium \citep{sciencewithcta}. The key science program consists of three different observational approaches: high quality spectra, long-term monitoring, and target-of-opportunity observations of flares.

The detailed $\gamma$-ray spectral studies with the improved sensitivity of CTA will enable us to detect the potential spectral features. Such features are expected due to $\gamma$-$\gamma$ absorption in the BLR or dusty torus. Even the detection of a cascade bump that should be the "smoking gun" of the hadronic processes is within the reach of CTA \citep[][see also the discussion in Section~\ref{hadronic}]{zech17}. In addition to these blazar intrinsic spectral features, the high-quality spectra will also be used to study extragalactic background light. The important aspect is that as there are several possible origins for the spectral features, it is crucial to observe a large sample of blazars from different sub-classes and in different redshifts. 

As discussed in Section~\ref{longterm}, long term VHE $\gamma$-ray light curves only exist for the three brightest VHE $\gamma$-ray blazars. Therefore, statistical properties of VHE $\gamma$-ray flaring behavior are largely unknown, as are also possible differences in flaring behavior of different sub-classes. In addition, the connection between events in lower frequencies and VHE $\gamma$-ray flares still lacks statistics (see Section~\ref{seedphotons}), as LSPs and FSRQs are detectable with short exposure times (if at all) only during flares. The long-term monitoring program of CTA, covering 15 AGN of different sub-classes, will be crucial to address these questions.  

A large fraction of VHE blazars that we know today have been detected during flares in lower energy bands, in particular in optical and HE $\gamma$-rays observed by {\it Fermi}-LAT. While this produces a certain observational bias to our VHE $\gamma$-ray light curves and AGN population studies, observing the flares is an absolutely mandatory part of VHE $\gamma$-ray observations, as even with CTA, many sources can only be detected during flares. In addition to a large number of flares from LSPs and FSRQs, also extending the AGN population to fainter classes, such as narrow-line Seyfert-1 galaxies is within the reach of CTA. The flares are important for studying the mechanism of fast (time scale of $<10$ minutes) variability. Also here, observing a large number of such flares is crucial, as from single flares from a single source, it is impossible to identify the mechanism that is causing the fast variability \citep[see also][]{bottcher19}. 

In addition, blazars will be targeted with proposal based programs. \citet{hassan17} demonstrated that according to the simulations, the number of blazars detectable by CTA will be almost an order of magnitude larger than with current generation of telescopes.

\subsection{The Square Kilometer Array}
The Square Kilometer Array (SKA)\footnote{\url{https://www.skatelescope.org}} is an array of radio telescopes to be located in South Africa and Australia. The name comes from its layout where thousands of telescopes will be placed to cover an area over a square kilometer to achieve a very large collecting area required for high sensitivity. In its first phase, with science commissioning anticipated to begin in 2022, the SKA will consist of $>200$ telescopes operating at 50\,MHz to 15\,GHz frequencies. In its second phase, the number of antennas will increase to thousands and the frequency coverage may be extended up to 24\,GHz. Already now, the pathfinder arrays MeerKAT \citep{meerkat16} in South Africa and ASKAP \citep{johnston07} in Australia are testing the instrument design and providing science data.

SKA will contribute to blazar science through its science goals on cosmic magnetism \citep{agudo15}, continuum surveys \citep{smolcic15}, and radio transients \citep{bignall15}. As discussed in \cite{bignall15}, with the SKA surveying the sky every day down to 100$\mu$Jy sensitivity, it will be possible to obtain timing information on a large number ($\sim10^6$) of objects with unprecedented sensitivity and cadence. This will allow studies of variability on multiple time scales. With the wide frequency coverage of SKA, it will also be possible to model the variations as a function of frequency, and gain insight into the emission processes. In addition to timing information, SKA will provide full-polarization spectra of thousands of objects, allowing the studies of their magnetic field composition \citep{agudo15}. As discussed in Section~\ref{Bfields}, full-polarization spectra, including sensitive circular polarization observations, are crucial for understanding the particle acceleration processes and the particle composition in jets. SKA data combined with other upcoming instruments, is thus expected to provide extremely interesting data for blazar science.

\subsection{IceCube-Gen2 and KM3Net}\label{icecube}

As discussed in Section \ref{multimessenger}, IceCube has started a new era in neutrino astronomy when astrophysical ultra-high-energy neutrinos were discovered in 2013 \citep{aartsen13}. This has given a strong motivation to build the next generation of neutrino observatories. A major upgrade to the IceCube detector at South Pole is planned{\footnote{\url{https://icecube.wisc.edu/science/beyond}}} as well as a major neutrino telescope KM3Net{\footnote{\url{http://www.km3net.org}}} to be built under the Mediterranean Sea. KM3Net will have a detector volume up to several cubic kilometres of clear sea water. These will provide a leap in sensitivity on the order of a magnitude, and therefore provide statistics to identify individual sources of neutrinos, such as blazars, with high confidence. Combined with multiwavelength observations, we should also be able to answer the open questions (see Section~\ref{hadronic}) about the hadronic models for blazars.  

\subsection{The Laser Interferometer Space Antenna}\label{lisa}
The Laser Interferometer Space Antenna (LISA) is a mission by the European Space Agency (ESA) to detect gravitational waves in the 0.1 to 100\,mHz range \citep{lisa17}. This frequency range is relevant to supermassive black holes, and provides a way to detect merging massive black holes in the centers of active galaxies and blazars. As discussed in \cite{lisa17}, the system should detect merging binaries in the mass range of $10^4-10^7$M$_\odot$ up to a redshift of 20. Higher mass binaries with masses of a few times $10^8$M$_\odot$ up to $10^9$M$_\odot$, more relevant to blazar scales, could still be detected in the nearby Universe at $z < 1$.

The system will consist of three identical spacecrafts that will be placed in a triangular formation with a 2.5 million km separation between the spacecrafts. Following the successful LISA pathfinder experiment \citep{armano16} where the technology was tested, the LISA mission was accepted by ESA with the launch expected to take place in 2034.


\section{Acknowledgments}
T. H. was supported by Academy of Finland projects 317383 and 320085. E. L. was supported by Academy of Finland projects 317636 and 320045.

\bibliographystyle{elsarticle-harv} 
\bibliography{references.bib}




%%\bibitem[ ()]{}

%%\end{thebibliography}
\end{document}